\documentclass[aps,prd,twocolumn,amsmath,amssymb,nofootinbib,superscriptaddress,10pt]{revtex4-1}

\usepackage{dsfont}
\usepackage{graphicx}
\usepackage{hyperref}
\usepackage[utf8]{inputenc}
\usepackage{makecell}
\usepackage{mathrsfs}
\usepackage{microtype}
\usepackage{placeins}
\usepackage{slashed}
\usepackage[caption=false]{subfig}
\usepackage{xcolor}

\newcommand{\cc}{\langle\psibar\psi\rangle}
\newcommand{\Dlatt}{D_\mathrm{latt}}
\newcommand{\Dnaive}{D_\mathrm{naive}}
\newcommand{\Dov}{D_\mathrm{ov}}
\newcommand{\Dslac}{D_\mathrm{SLAC}}
\newcommand{\Dw}{D_W}
\newcommand{\g}{\gamma_5}
\newcommand{\ghat}{\hat{\gamma}_5}
\newcommand{\id}{\mathds{1}}
\newcommand{\ii}{\mathrm{i}}
\newcommand{\Nf}{N_{\mathrm{f}}}
\newcommand{\Ns}{N_{\mathrm{s}}}
\newcommand{\Nt}{N_{\mathrm{t}}}
\newcommand{\sig}{\langle\vert\bar{\sigma}\vert\rangle}
\newcommand{\psibar}{\bar{\psi}}
\newcommand{\Tc}{T_{\mathrm{c}}}
\newcommand{\Z}{\mathbb{Z}}

\newcommand{\sref}[1]{Sec.~\ref{#1}}
\newcommand{\aref}[1]{App.~\ref{#1}}
\newcommand{\tref}[1]{Tab.~\ref{#1}}
\newcommand{\fref}[1]{Fig.~\ref{#1}}

\renewcommand{\eqref}[1]{Eq.~(\ref{#1})}
\renewcommand{\L}{\mathcal{L}}

\DeclareMathOperator{\tr}{tr}

\graphicspath{{./figures/}}

\begin{document}

	\title{Magnetic catalysis in the (2+1)-dimensional Gross-Neveu model}
	
	\author{Julian J. Lenz}
	\email{j.j.lenz@swansea.ac.uk}
	\affiliation{Theoretisch-Physikalisches Institut, Friedrich-Schiller-Universität Jena,  D-07743 
	Jena, Germany}
	\affiliation{Swansea Academy of Advanced Computing, Swansea University, Fabian Way, SA1 8EN, 
	Swansea, Wales, UK}
	\author{Michael Mandl}
	\email{michael.mandl@uni-jena.de}
	\affiliation{Theoretisch-Physikalisches Institut, Friedrich-Schiller-Universität Jena,  D-07743 
	Jena, Germany}
	\author{Andreas Wipf}
	\email{wipf@tpi.uni-jena.de}		
	\affiliation{Theoretisch-Physikalisches Institut, Friedrich-Schiller-Universität Jena,  D-07743 
	Jena, Germany}
	
	\date{\today}
	
	\begin{abstract}	
		We study the Gross-Neveu model in $2+1$ dimensions in an external magnetic field $B$. We 
		first summarize known mean-field results, obtained in the limit of large flavor number $\Nf$, 
		before presenting lattice results using the overlap discretization to study one reducible 
		fermion flavor, $\Nf=1$. Our findings indicate that the magnetic catalysis phenomenon, i.e., 
		an increase of the chiral condensate with the magnetic field, persists beyond the mean-field 
		limit for temperatures below the chiral phase transition and that the critical
		temperature grows with increasing magnetic field. This is in contrast to the situation in QCD, where the
    broken phase shrinks with increasing $B$ while the	condensate exhibits a
    non-monotonic $B$-dependence close to the chiral crossover, and we comment on this discrepancy. We do not find
    any trace of inhomogeneous phases induced by the magnetic field.
  \end{abstract}
	\maketitle

	\section{Introduction}\label{sec:introduction}
	In recent years the study of strongly interacting quantum field theories exposed to external
electromagnetic fields has received a significant amount of attention in the high-energy physics
community. This is due to the fact that magnetic fields are believed to play an important role 
in a plethora of physical processes, such as heavy-ion collisions 
\cite{RM76,KMW08,Tuc13,BPT13,XSZ20}, the strong interactions within neutron 
stars \cite{LS91,BBG95,HL06r,IPH21r}, and at several stages of the early Universe 
\cite{Vac91,EO93,BBM96,HE98} -- see \cite{MS15r} for an extensive review. 

Quantum Chromodynamics (QCD) is the theoretical framework underlying the strong interactions 
and -- as such -- describes the aforementioned phenomena. Since 
QCD cannot be studied using perturbation theory in the parameter regime of 
interest, one has to resort to non-perturbative methods, of which lattice quantum field theory 
is the most reliable one. Lattice simulations, however, suffer from the infamous complex-action 
problem, rendering the use of conventional Monte-Carlo methods impossible at finite density.

Needless to say, there are countless attempts aiming at circumventing the complex-action 
problem (see, e.g., \cite{GL16}), but none of them has fully solved it within finite-density QCD. In 
this work we employ a different approach altogether, using a low-dimensional toy model, the 
Gross-Neveu (GN) model \cite{GN74}, as an effective description of QCD. This is motivated by the 
fact that the GN model shares a number of important features with QCD, such as chiral symmetry 
and its spontaneous breakdown, or (in $3$ dimensions or less) renormalizability \cite{RWP89,RWP91}. 

It should be mentioned that there exist more realistic models, bearing a closer 
similarity to QCD than the one considered in this work, for instance models of the 
Nambu--Jona-Lasinio (NJL) \cite{NJL61} or quark-meson (see, e.g., \cite{GL60}) type. Still, the 
simplicity of the GN model merits its use as a starting point for the search for a description of QCD 
using effective models, which may then be expanded upon. 

Furthermore, we mention that the GN model and variants thereof are also
interesting from a condensed-matter perspective as they have been used
successfully to describe certain one-dimensional and planar materials, such as
polymers \cite{CB81,CB82,CM94,Cal11}, graphene \cite{DS08,JHS09,EKK16,EB19}, and
high-temperature superconductors \cite{Liu99,ZKK01,Thi03}.  One should, however,
take some care in translating the results because the mapping of physical
(non-relativistic) degrees of freedom to the field-theoretical description with
emergent Lorentz invariance is not always straightforward -- see \cite{Thi06} 
for one example.

Four-Fermi theories, including GN-type models, have been extensively studied in the
literature with a variety of methods. A first -- and often quite reasonable --
approximation is given by mean-field treatments, which become exact at infinitely
large flavor numbers $\Nf$ and can be systematically improved by expanding 
in orders of $1/\Nf$.
Since we use the mean-field behavior as a guideline and as an important comparison for
our lattice investigation, we summarize the most relevant known results for our
system of interest in the following.
 
Early attempts to study the GN
model in three space-time dimensions including an external magnetic field were
made in \cite{Kli91} and extended to finite temperature in \cite{Kli92,Kli92_2}.
It was found that the magnetic field is a strong catalyst of chiral
symmetry breaking, enhancing the chiral condensate at both zero and non-zero
temperature. This effect, termed \emph{magnetic catalysis}, was explained in
\cite{GMS94} to be caused by a dimensional reduction due to the
applied field, similar to the effect of the Fermi surface in superconductivity
-- see also the reviews \cite{Sho13r,ANT16r}. 

The goal of this work is to investigate whether the magnetic catalysis in the 
GN model is merely an artifact of the large\,-$\Nf$ limit, where quantum 
fluctuations are suppressed, or is present in the full theory at finite flavor 
number as well. Work in this direction has already been performed using methods 
superior to the mean-field approximation, such as the functional renormalization 
group \cite{SG12} or optimized perturbation theory \cite{KPR13}, both supporting 
the presence of magnetic catalysis also at finite $\Nf$. However, we are not 
aware of any lattice studies of the GN model in an external electromagnetic 
field to be found in the literature at the time of writing, and we attempt to fill this 
gap. To this end, we perform extensive lattice simulations using overlap 
fermions at both zero and non-zero temperature and for vanishing chemical 
potential. The finite-density case will be studied in detail in an upcoming 
publication.

We note that -- to the best of our knowledge -- this is the first lattice Monte-Carlo 
simulation of GN-type models that uses overlap fermions 
(although theoretical considerations already exist in the literature
\cite{IN00}).
Thus, we put considerable effort into working out the technical details and
intricacies involved.  However, we
decided that they are better suited to be part of another planned publication
with a more technical and analytical focus.

As we value the reproducibility of our results according to the FAIR 
Guiding Principles \cite{FAIR16} (see \cite{ABLP22} for a recent study about
its status in our community), we provide access to our 
simulation results in \cite{data}. Furthermore, the scripts used to perform 
our data analyses can be found in \cite{code}.

This work is structured as follows: In \sref{sec:analytical} we introduce the GN model in an 
external magnetic field and discuss its large\,-$\Nf$ limit. Our lattice formalism is outlined in 
\sref{sec:lattice}, and our results are presented in \sref{sec:results}. Finally, we discuss and 
critically analyze our findings and put them into perspective with respect to known QCD results in 
\sref{sec:discussion}.

	\section{Analytical results}\label{sec:analytical}
	The GN model in its most basic form is given by the Lagrangian \cite{GN74}
\begin{align}\label{eq:lagrangian_4f}
	\L = \psibar\ii\slashed{\partial}\psi + \frac{g^2}{2\Nf}(\psibar\psi)^2\;,
\end{align}
featuring $\Nf$ flavors of fermionic fields, collected implicitly in the tuple $\psi$ and self-interacting via a scalar-scalar channel
with coupling constant $g^2$. The sum over flavors is implied in (\ref{eq:lagrangian_4f}). 

In
order to bring the model into a form amenable to our mean-field treatment as well as to lattice
simulations, one introduces an auxiliary scalar field $\sigma$ by means of a Hubbard-Stratonovich
transformation. The semi-bosonized, but fully equivalent, theory then reads
\begin{align}\label{eq:lagrangian}
	\L_\sigma = \ii\psibar(\slashed{\partial}+\ii e\slashed{A}+\sigma)\psi + \frac{\Nf}{2g^2}
	\sigma^2\;,
\end{align}
where we have furthermore coupled the fermions to an external vector field $A_\mu$ and $e$ denotes the 
elementary electric charge. 

For the remainder of this work we shall be concerned with a $(2+1)$-dimensional
space-time and four-component spinor fields transforming in a reducible
representation of the Dirac algebra \cite{Pis84}.
This allows one to introduce a ``fifth''\footnote{More precisely, the reducible
representation of the Clifford algebra contains two linearly independent matrices anti-commuting with
all other elements. The respective $\Z_2$ symmetries they generate, however, are
not independent. This is because the product of these matrices is non-trivial
and commutes with all other elements of the Clifford algebra, giving rise to a
further $U(\Nf)$ symmetry which relates the (seemingly) independent $\Z_2$ factors. This
$U(\Nf)$, however, is irrelevant for us since it persists in the presence of a
chiral condensate.  Further information can be found, e.g., in
\cite{GJ10}.\label{fn:gamma5}} gamma matrix $\g$, anti-commuting with all other
$ \gamma_\mu$. The Lagrangian (\ref{eq:lagrangian}) is then invariant under a
discrete $\Z_2$ chiral transformation:
\begin{align}\label{eq:chiral_symmetry_continuum}
	\psi\rightarrow\g\psi\;, \quad \psibar\rightarrow-\psibar\g\;, \quad \sigma
	\rightarrow-\sigma\;.
\end{align}
It is this chiral symmetry and its spontaneous breaking that will be our main concern in this 
work. 

As an order parameter for chiral symmetry breaking we consider the fermion condensate $\cc$, which is 
proportional to the expectation value of the auxiliary field $\sigma$ by means of a Dyson-Schwinger 
equation:
\begin{align}\label{eq:ward_identity}
	\cc = \frac{\ii\Nf}{g^2}\langle\sigma\rangle\;.
\end{align}
In the limit of an infinite flavor number, the path integral defining the partition function of 
the model,
\begin{align}\label{eq:partition_function}
	Z = \int\mathcal{D}\psibar\mathcal{D}\psi\mathcal{D}\sigma e^{-S[\psibar,\psi,\sigma]}\;, \ 
  \ S[\psibar,\psi,\sigma] = \int\! \mathrm{d}^3x\,\L_\sigma\;,
\end{align}
is, after integrating out the fermions, reduced to the problem of minimizing the effective 
potential
\begin{align}\label{eq:effective_potential}
	V_\mathrm{eff} = \frac{V}{2g^2}\sigma^2-\ln\det D\;,
\end{align}
where we have assumed $\sigma$ to be homogeneous in space and time, $V$ denotes the space-time 
volume and $D$ is the Dirac operator
\begin{align}
	D = \slashed{\partial} + \ii e\slashed{A} + \sigma\;.
\end{align}

\begin{figure}[t]
	\includegraphics[width=\linewidth]{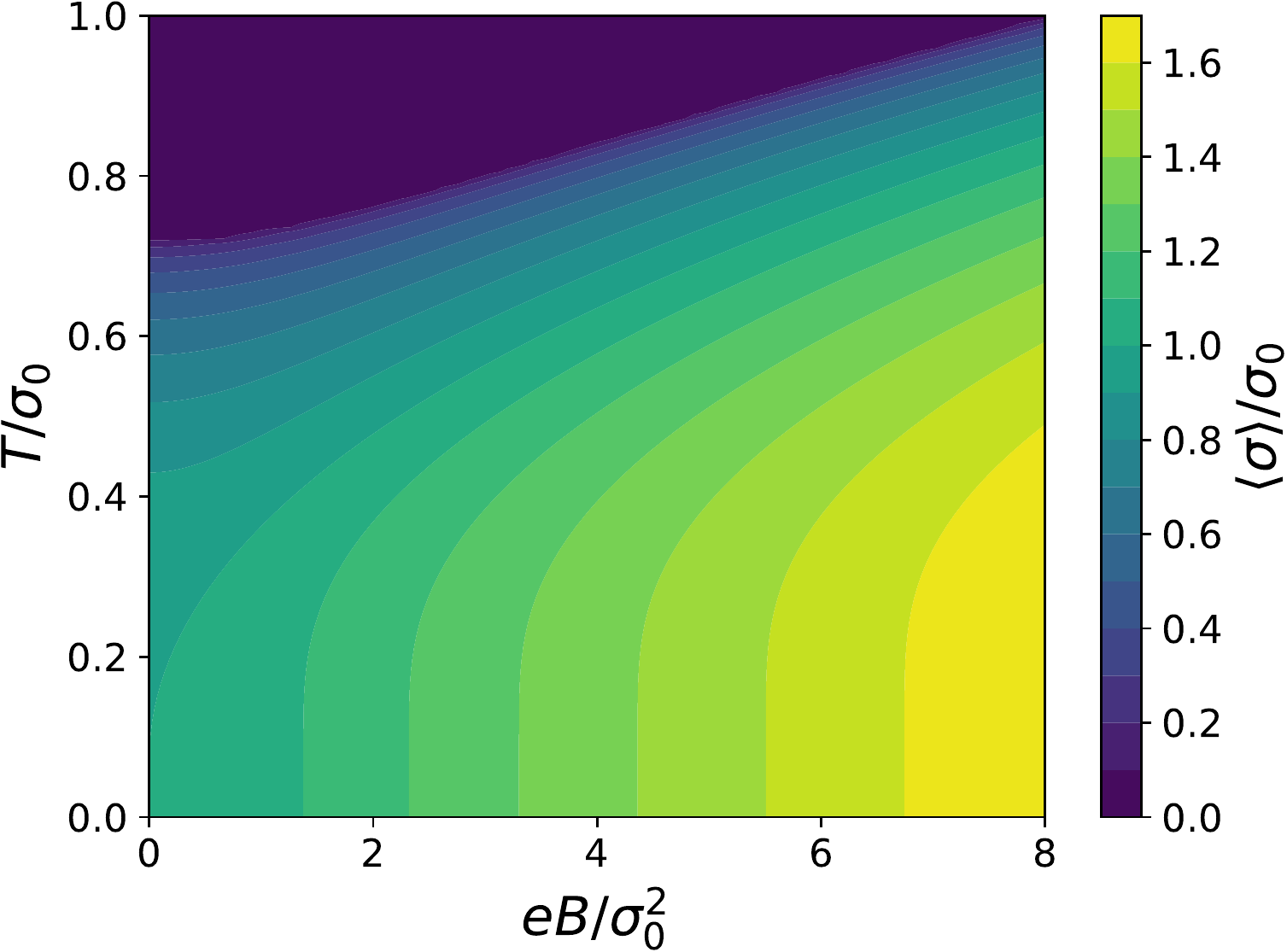}
	\caption{Large\,-$\Nf$ phase diagram in the $(B,T)$ plane. $\sigma_0$ denotes the value of 
	$\langle\sigma\rangle$ at vanishing $T$ and $B$.}
	\label{fig:pd_large_N}
\end{figure}

For $A_\mu$ describing a constant and homogeneous (electro)magnetic field
$B$ and, without loss of generality, assuming $B>0$, one finds the following
effective potential density \cite{GMS95}:
\begin{align}\label{eq:veff_closed_form}
\notag	\frac{V_\mathrm{eff}}{V} 
		= -&\frac{\sigma^2}{2\pi}\sigma_0
		 - \frac{\sqrt{2}}{\pi}(eB)^{3/2}\zeta_H\left(-\frac{1}{2}, \frac{\sigma^2}{2eB}\right) 
	     + \frac{\vert\sigma\vert eB}{2\pi} \\
	    - \frac{eB}{\pi\beta}&\sum_{l=0}^{\infty}d_l\ln\left(1+\exp\left(-\beta\sqrt{\sigma^2+2eBl}
	    \right)\right)\;,
\end{align}
where $\sigma_0>0$ denotes the minimum of $V_\mathrm{eff}$ at vanishing temperature and magnetic 
field, $\zeta_H$ is the Hurwitz zeta function, $\beta=1/T$ 
denotes the inverse temperature, and the last term is a sum over Landau levels $l$ with degeneracies 
$d_l=2-\delta_{l0}$. Note that we work in the strong-coupling regime, where chiral symmetry is 
spontaneously broken at vanishing $T$ and $B$, i.e., $\sigma_0\neq0$, which is not the case for weak 
couplings. Remarkably, the volume-dependence of $V_\mathrm{eff}$ is contained only in the 
discretization of $eB$ in a finite volume, see \eqref{eq:flux_quantization} below. For a derivation 
of \eqref{eq:veff_closed_form}, see \aref{app:large_N}. The 
global minima $\langle\sigma\rangle$ of the effective potential for different temperatures and 
magnetic field 
strengths determine the mean-field phase structure of the GN model, which we show in 
\fref{fig:pd_large_N}.

Evidently, chiral symmetry is spontaneously broken (i.e., $\langle\sigma\rangle\neq0$) for low 
temperatures and $B=0$. The magnetic field then enhances this breaking even further, causing the 
chiral condensate to increase. This is the magnetic catalysis phenomenon 
mentioned in the Introduction. We furthermore observe that the critical temperature $T_c(B)$, beyond 
which chiral symmetry is restored (i.e., $\langle\sigma\rangle=0$), increases monotonically with $B$, and thus, the region of broken symmetry grows with the magnetic field. 

We remark at this point that the magnetic-field-induced dimensional reduction down to one 
space-time dimension, found in \cite{GMS94,GMS95,GMS96} to be responsible for magnetic catalysis, is 
not in conflict with the no-go theorem prohibiting the existence of phases in one dimension \cite{LL80B} (not to be confused with the 
Coleman-Hohenberg-Mermin-Wagner theorem \cite{MW66,Hoh67,Col73} preventing the spontaneous 
breaking of \emph{continuous} symmetries in \emph{two} dimensions). This is due to the fact that the 
chiral condensate itself is electrically neutral and, thus, unaffected by the dimensional reduction. For a similar argument in the $U(2)$-symmetric NJL model, see \cite{GMS95}.

It is the main purpose of this work to shed light on the fate of the results presented in this 
section when going beyond the mean-field limit, i.e., when considering a finite number of 
fermionic flavors $\Nf$ and lifting the restriction of homogeneity on $\sigma$.

	\section{Lattice setup}\label{sec:lattice}
	\subsection{Discretization}\label{sec:lattice_discretization}

We intend to study the theory with Lagrangian (\ref{eq:lagrangian}) on a three-dimensional lattice $\Lambda$ with 
$N_\mu$ lattice points in the $x_\mu$-direction ($\mu=0,1,2$) and an isotropic lattice constant 
$a$. 
For the entirety of this work we shall always consider $N_1$ and $N_2$ to be equal, $N_1=N_2=:\Ns$, 
such that the physical lattice extent in each spatial direction is given by $L=a \Ns$. 
Furthermore, we introduce $\Nt:=N_0$ to denote the number of lattice points in the (Euclidean) time 
direction, such that the inverse temperature reads $\beta=a\Nt$. We then denote the space-time volume 
as $V=L^2\beta$. The bosonic field $\sigma$ obeys periodic boundary conditions in all directions, 
while the fermions are periodic in space and anti-periodic in time.

The question of which lattice discretization to use for fermions is a non-trivial
one. Studies of QCD in background magnetic fields mainly
rely on the use of staggered fermions \cite{DMS10,DN11,BBE12_2,BBE12} (with a few authors employing overlap
fermions as well \cite{BBC14}). However, it has become clear that staggered
fermions can be problematic in asymptotically safe theories \cite{HRW18p,WSW17,LWW19,HMW20,Han19}, of 
which (\ref{eq:lagrangian}) is an example. 
Moreover, since we
are interested in studying chiral symmetry, we refrain from using Wilson fermions, and since we prefer
to avoid the fermion doubling problem, we cannot use the naive discretization for $\Nf<8$ either.
Finally, even though in previous works \cite{LWW19,LPW20,LPW20_1,LMW22} the SLAC derivative 
\cite{DWY76_1,DWY76_2} has
proven to be the best-suited discretization for studying GN-like theories on the lattice, it fails
when naively applied to theories with gauge symmetry \cite{KS79}. As a matter of fact, it is not obvious how
to properly formulate the GN model in a magnetic field with SLAC fermions in a gauge-invariant way in 
the first place.
We nevertheless discuss this issue further and provide a more detailed comparison between different
possible discretizations in \aref{app:discretizations}.

We are left with the choice of employing Ginsparg-Wilson fermions \cite{GW82}, which have 
ideal chiral properties but come with a significantly increased cost due to their 
non-ultralocality \cite{Hor98}. For our lattice studies we use Neuberger's formulation 
\cite{Neu98} of the overlap operator \cite{NN93,NN94}, reading\footnote{We remark that this 
expression does not make use of $\g$ and could thus be used in an irreducible representation 
of gamma matrices in $(2+1)$ dimensions as well \cite{KN98,BN01}.}
\begin{align}\label{eq:overlap_massless}
	\Dov = \frac{1}{a}\left(\id+A/\sqrt{A^\dagger A}\right)\;.
\end{align}
Here, the kernel $A$ is given by the Wilson operator $\Dw$ with a negative mass 
parameter $m=-1$:
\begin{align}
	A &= a\Dw-\id\;,\label{eq:wilson_kernel}\\
	\Dw = \frac{1}{2}\big[\gamma_\mu\big(&\nabla_\mu^*+\nabla_\mu\big) - 
	a\nabla_\mu^*\nabla_\mu\big]\;,\label{eq:wilson}
\end{align}
where the action of the covariant forward and backward difference operators on $\psi(x)$ is 
defined as
\begin{align}\label{eq:covariant_difference}
	\begin{aligned}
		\nabla_\mu\psi(x) &= \frac{1}{a}\left[U_\mu(x)\psi\left(x+a\hat{\mu}\right)-\psi(x)
		\right]\;,\\
		\nabla_\mu^*\psi(x) &= \frac{1}{a}\left[\psi(x)-U_\mu^\dagger(x-a\hat{\mu})\psi\left(x-a
		\hat{\mu}\right)\right]\;.
	\end{aligned}
\end{align}
In (\ref{eq:covariant_difference}), $\hat{\mu}$ denotes the unit vector in the $x_\mu$-direction, and
\begin{align}\label{eq:gauge_links}
	U_\mu(x)=e^{\ii a e A_\mu(x)}
\end{align}
are $U(1)$ link variables.

Guided by the Lagrangian (\ref{eq:lagrangian}), where the Yukawa term $\sigma\psibar\psi$ would
reduce to a fermionic mass term if $\sigma$ was constant, one can introduce the scalar
field into the overlap formalism by the definition \cite{IN00}
\begin{align}\label{eq:overlap}
	D = \Dov + \sigma\left(\id-\frac{a}{2}\Dov\right)
\end{align}
for the full Dirac operator.\footnote{Note that this definition differs from the one given 
in \cite{ISU02}.} For constant $\sigma$ the second term in \eqref{eq:overlap} is just a mass 
term in Ginsparg-Wilson language \cite{Cha99} (see \cite{VTK00} for a similar argument in the 
domain-wall formalism). By this definition one ensures that the identity 
(\ref{eq:ward_identity}), relating the expectation value of $\sigma$ to the chiral condensate, 
is preserved\footnote{The fact that a factor of $\ii$ is missing when comparing \eqref{eq:chiral_condensate} to \eqref{eq:ward_identity} is purely conventional and has no influence on any of the results or their interpretation.} on the lattice, i.e.,
\begin{align}\label{eq:chiral_condensate}
	-\frac{\Nf}{g^2}\langle\sigma\rangle = \cc_\mathrm{ov} := \left\langle\psibar\left(\id-\frac{a}{2}\Dov
	\right)\psi
	\right\rangle\;,
\end{align}
facilitating the numerical study of chiral symmetry breaking considerably.
The full action of our lattice theory thus reads
\begin{align}\label{eq:lattice_action}
	S = \psibar D\psi + \frac{\Nf}{2g^2}\sigma^2\;,
\end{align}
where summation over space-time and internal indices is implied. 

The discrete symmetry 
(\ref{eq:chiral_symmetry_continuum}) of the continuum theory has an exact lattice counterpart 
in the overlap formalism, much like the case for theories with the more common $U(1)$ 
chiral symmetry \cite{Lue98}. Namely, introducing $\ghat=\g(\id-a\Dov)$, we find that the action 
(\ref{eq:lattice_action}) is invariant under
\begin{align}\label{eq:chiral_symmetry_lattice}
	\psi\rightarrow\ghat\psi\;, \quad \psibar\rightarrow-\psibar\g\;, \quad \sigma
	\rightarrow -\sigma\;,
\end{align}
by using the Ginsparg-Wilson relation \cite{GW82}
\begin{align}\label{eq:ginsparg_wilson}
	\left\{\Dov,\g\right\} = a\Dov\g\Dov\;.
\end{align}
It should be noted that the additional symmetries in the continuum theory that arise due to 
ambiguity in the choice of the ``fifth'' gamma matrix (see footnote \ref{fn:gamma5}) can also be 
exactly translated to the lattice \cite{Han15}, but are not of interest in this work.

From \aref{app:discretizations} we know that (massive) overlap 
fermions suffer from discretization effects that quantitatively change the chiral
condensate in a theory of free fermions. Thus, one should investigate the interacting
theory with a particular emphasis on its behavior towards the continuum limit to see
if the discretization effects persist.

\subsection{Magnetic field on the lattice}\label{sec:lattice_magnetic_fields}

It is well known that the magnetic flux through a torus with an area $L^2$, orthogonal to an 
applied magnetic field $B$, is necessarily quantized \cite{tHo81,SW92}. One finds the 
following quantization condition for the magnetic field:
\begin{align}\label{eq:flux_quantization}
	eB = \frac{2\pi}{L^2}b\;, \quad b\in\mathbb{Z}\;.
\end{align}

\begin{figure}[t]
	\centering
	\includegraphics[scale=0.9]{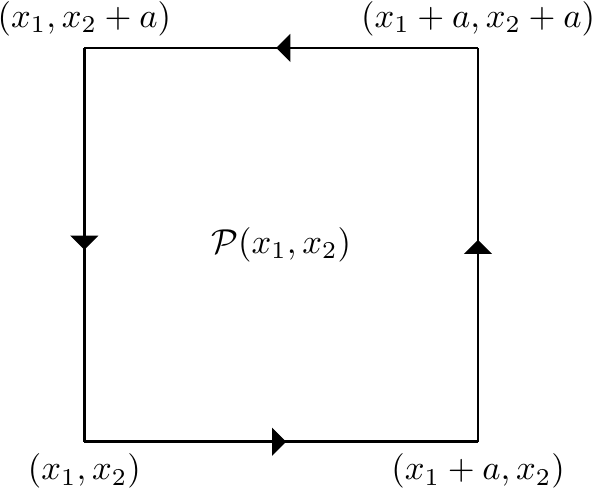}
	\caption{Plaquette at position $(x_1,x_2)$ in the spatial plane.}
	\label{fig:plaquette}
\end{figure}

Let us now outline how to implement an external magnetic field perpendicular to the spatial 
plane using the gauge links (\ref{eq:gauge_links}) in our lattice formulation 
(\ref{eq:lattice_action}). In the continuum one could represent such a magnetic field by, e.g., 
the following choice of vector potential:
\begin{align}\label{eq:continuum_vector_potential}
	A_0(x) = 0\;, \quad A_1(x) = 0\;, \quad A_2(x) = Bx_1\;.
\end{align}
On a lattice with periodic boundary conditions, however, this definition does not lead to a 
constant magnetic flux 
\begin{align}\label{eq:magnetic_flux}
	\Phi_\mathcal{P} = \oint_\mathcal{P} A_\mu ds_\mu
\end{align}
through every lattice plaquette $\mathcal{P}(x_1, x_2)$ in the spatial plane at position 
$(x_1, x_2)$ -- see \fref{fig:plaquette} for the definition of such a plaquette and the 
integration path in \eqref{eq:magnetic_flux}.
In fact, one finds
\begin{align}
	\begin{aligned}
		\Phi_\mathcal{P} = \begin{cases} a^2B \quad &\textnormal{if} \quad 0 \leq x_2 < L-a \\
		                                 a^2B -aBL \quad &\textnormal{if} \quad x_2=L-a
						   \end{cases}\;,
	\end{aligned}
\end{align}
i.e., the flux through the lattice boundary in the $x_2$-direction is large and opposite to the flux 
through the bulk, such that the total magnetic flux through the lattice vanishes:
\begin{align}
	\Phi_\mathrm{tot} = \sum_\mathcal{P}\Phi_\mathcal{P} = 0\;.
\end{align}

The solution is to introduce correction terms in $A_\mu$ on the lattice boundary in a way that shifts all the negative 
(assuming $B>0$) flux to the single plaquette at the combined boundary $x_1=x_2=L-a$. This can 
be achieved by the following definition \cite{RBK11}:
\begin{align}\label{eq:lattice_vector_potential}
	\ A_1(x) = -\frac{BL}{a} x_2 \delta_{x_1, L-a}\;, \ A_2(x) = Bx_1\;,
\end{align}
with $A_0$ set to zero. The flux through $\mathcal{P}(L-a,L-a)$ is now given by 
\begin{align}\label{eq:boundary_flux}
	\Phi_\mathcal{P}\vert_{x_1=x_2=L-a} = a^2B-BL^2= a^2B-\frac{2\pi}{e} b\;,
\end{align}
where we have used (\ref{eq:flux_quantization}), and $\Phi_\mathcal{P}=a^2B$ everywhere else. 
Since in our lattice formulation $A_\mu$ only appears in exponentials due to 
(\ref{eq:gauge_links}), the only way $\Phi_\mathcal{P}$ contributes is via the plaquette terms 
\begin{align}
	U_{\mu\nu}(x) = U_\mu(x) U_\nu(x+a\hat{\mu}) U_\mu^\dagger(x+a\hat{\nu}) U_\nu^\dagger(x)\;,
\end{align}
as we have
\begin{align}
	U_{12}(x) = e^{\ii e\Phi_\mathcal{P}}\;.
\end{align}
In this last expression the term proportional to $2\pi$ in (\ref{eq:boundary_flux}) cancels out. We 
thus end up with a situation that is physically indistinguishable from one with a constant 
magnetic flux $\Phi_\mathcal{P}=a^2B$ through every plaquette and a non-vanishing total flux
\begin{align}
	\Phi_\mathrm{tot} = \frac{2\pi}{e} b\;,
\end{align}
as desired.

We therefore use the following definition of the $U(1)$ gauge links $U_\mu(x)$ in
(\ref{eq:gauge_links}), entering the Wilson operator (\ref{eq:wilson}) via 
(\ref{eq:covariant_difference}):
\begin{align}\label{eq:gauge_links_magnetic_field}
	\begin{aligned}
		U_0(x) &= 1\;, \\
		U_1(x) &= \begin{cases}e^{- 2 \pi \ii b x_2 / L} \ &\textnormal{if}\ x_1 =L-a \\
	        	  	               1 \quad &\textnormal{else}
	    		      \end{cases}\;, \\
		U_2(x) &= e^{2 \pi \ii a b x_1 / L^2}\;.
	\end{aligned}
\end{align}
We see that the compactness of the gauge links introduces a periodicity in the magnetic field 
and hence an effective upper bound for the flux quantum number $b$, i.e.,
\begin{align}
	0 \leq b \leq \Ns^2\;.
\end{align}
In practice, one restricts $b$ even further in order to avoid discretization artifacts 
\cite{DN11,BE11} and we shall do the same in this work, performing simulations only up to $b 
\lesssim \Ns^2/16$.

\subsection{Computational details}

Our lattice setup of the GN model in a magnetic field, using the overlap Dirac operator 
(\ref{eq:overlap}), has a significant computational advantage compared to the use of overlap 
fermions in gauge theories. This is due to the fact that in our case the gauge links are not 
dynamical, depending only on the constant magnetic field. This allows for an \emph{exact} 
computation of the massless overlap operator $\Dov$ in (\ref{eq:overlap_massless}) that we perform
once,
at the beginning of a simulation. We then re-use $\Dov$ in every update step for the now 
straightforward computation of the full operator (\ref{eq:overlap}). Needless to say, computing 
the overlap operator exactly, i.e., without using approximations (see, e.g., \cite{EFL02}),
would be unthinkable in realistic QCD simulations.

For this work we have performed simulations at various temperatures and magnetic fields 
using a standard rHMC algorithm. We change the temperature by varying $\Nt$ at constant $\Ns$, 
and we study different lattice spacings by changing the coupling $g^2$ while simultaneously 
adjusting $\Ns$ such that the physical lattice volume remains constant. We 
furthermore approach larger physical volumes by increasing $\Ns$ at fixed $g^2$. Finally, we 
mention that our theory does not suffer from a complex-action problem, as is shown in 
\aref{app:sign_problem}.

\subsection{Observables}\label{sec:observables}

As the order parameter for chiral symmetry breaking, the main observable of interest is the 
chiral condensate $\langle\sigma\rangle$ in (\ref{eq:chiral_condensate}). Assuming an ergodic 
simulation algorithm, however, this quantity will average to zero. This is because the 
effective potential of the GN model is known to exhibit two equivalent minima in the 
spontaneously broken phase, differing only in the sign of $\sigma$, hence leading to a 
cancellation between those minima. In order to avoid this cancellation, we thus use the 
quantity 
\begin{align}\label{eq:order_parameter}
	\sig\;, \quad \textnormal{with}\ \bar{\sigma} = \frac{1}{V}\sum_{x\in\Lambda}\sigma(x)
\end{align}
as an order parameter instead \cite{KS01}. Here, the sum runs over the whole lattice and 
$\langle\,\cdot\,\rangle$ denotes the Monte-Carlo average. While $\sig$ approaches $\pm\langle\sigma\rangle$ in the infinite-volume limit, one should keep in 
mind that on finite volumes $\sig$ will never be zero exactly, even when chiral symmetry is intact, 
which complicates the study of phase transitions. For this reason, $\sig$ should -- strictly speaking -- not be referred to as an order parameter. However, for the sake of convenience we will still do so in the following.

In order to find the critical temperature $\Tc$, corresponding to the phase transition between the 
two respective regions of spontaneously broken and restored chiral symmetry, we study the chiral 
susceptibility, defined as\footnote{The
factor of $V$ is compensated by the use of space-time-averaged
quantities in the expectation values such that $\chi$ is an intensive quantity,
as it should be.}
\begin{align}\label{eq:susceptibility}
	\chi = V\left(\langle\bar{\sigma}^2\rangle - \sig^2\right)\;,
\end{align}
as a function of $T$. Approaching a second-order phase transition, $\chi(T)$ diverges
rationally. This behavior is washed out by finite-volume corrections and we
expect to find a sharp but smooth peak close to the transition temperature that
monotonically grows, sharpens and moves towards the latter \cite{Jan08}.

At this point we should mention that the introduction of an additional length
scale and some form of imbalance in fermionic theories might induce
spatial inhomogeneities in the system \cite{RBD14}. While this is most prominently observed
in mean-field treatments at finite density \cite{Thi06,BC16}, it could also
apply to external magnetic fields. In fact, it is known that in $3+1$ dimensions 
magnetic fields can favor inhomogeneous condensates at finite density when they
would be disfavored at $B=0$ \cite{FZK10,BSE10,TNK15,BC16}. This 
can be understood by recalling the dimensional reduction induced by the 
magnetic field \cite{GMS94} and the fact that inhomogeneous 
phases are more abundant in lower dimensions. Of course, our situation is 
qualitatively different because in our $(2+1)$-dimensional setup the dimensional
reduction in a strong magnetic field leaves us with no spatial dimension at all
(and we do not expect inhomogeneities in the temporal direction in equilibrium).

Since in $2+1$ dimensions there is no conclusive evidence for the existence of 
inhomogeneous structures beyond mean-field (as compared to the ($1+1$)-dimensional 
case \cite{LPW20,LPW20_1,LMW22,HN21p}) and there even exist some negative mean-field results 
\cite{BKW21,PWW22,WP22}, such inhomogeneities are not the focus of this work. Nonetheless, 
since the previous studies did not take into account the influence of 
magnetic fields, we also investigate whether 
an external magnetic field can induce inhomogeneities in $2+1$ dimensions at
zero density. To this end, we follow \cite{LPW20} by introducing the spatial correlation 
function
\begin{align}\label{eq:spatial_correlator}
	\begin{aligned}
		C(x_1, &x_2) = \\
		\frac{1}{\Ns^2\Nt}&\sum_{x'\in\Lambda}\left\langle\sigma(x'_0, x_1, x_2)\sigma(x'_0, 
		x_1+x'_1,x_2+x'_2)\right\rangle\;.
	\end{aligned}
\end{align}
As has been outlined in \cite{LPW20}, this correlator should capture any 
inhomogeneities if they exist.

\subsection{Scale setting}\label{sec:scale_setting}

We set the scale via the order parameter at vanishing magnetic field and the lowest 
temperature considered, $T_0\approx0$:
\begin{align}\label{eq:scale}
	\sigma_0 := \sig_{B=0,\, T=T_0}\;.
\end{align}
We keep $T_0$ constant as we approach the infinite-volume ($L^2\rightarrow\infty$ at fixed $a$) and 
continuum ($a\rightarrow0$ at fixed $L^2$) limits, respectively. However, in order to ensure reasonably 
low scale-setting temperatures at an affordable computational cost, we consider two different $T_0$ 
corresponding to the two different limits. 

For a detailed list of the parameters we have performed simulations for as well as their 
corresponding $\sigma_0$ and $T_0$, we refer to \tref{tab:parameters} in \aref{app:parameters}. In 
\aref{app:parameters} we also give a brief description of how the error estimates presented in this work 
are obtained.

	\section{Results}\label{sec:results}
	\begin{figure}[t]
	\includegraphics[width=0.95\linewidth]{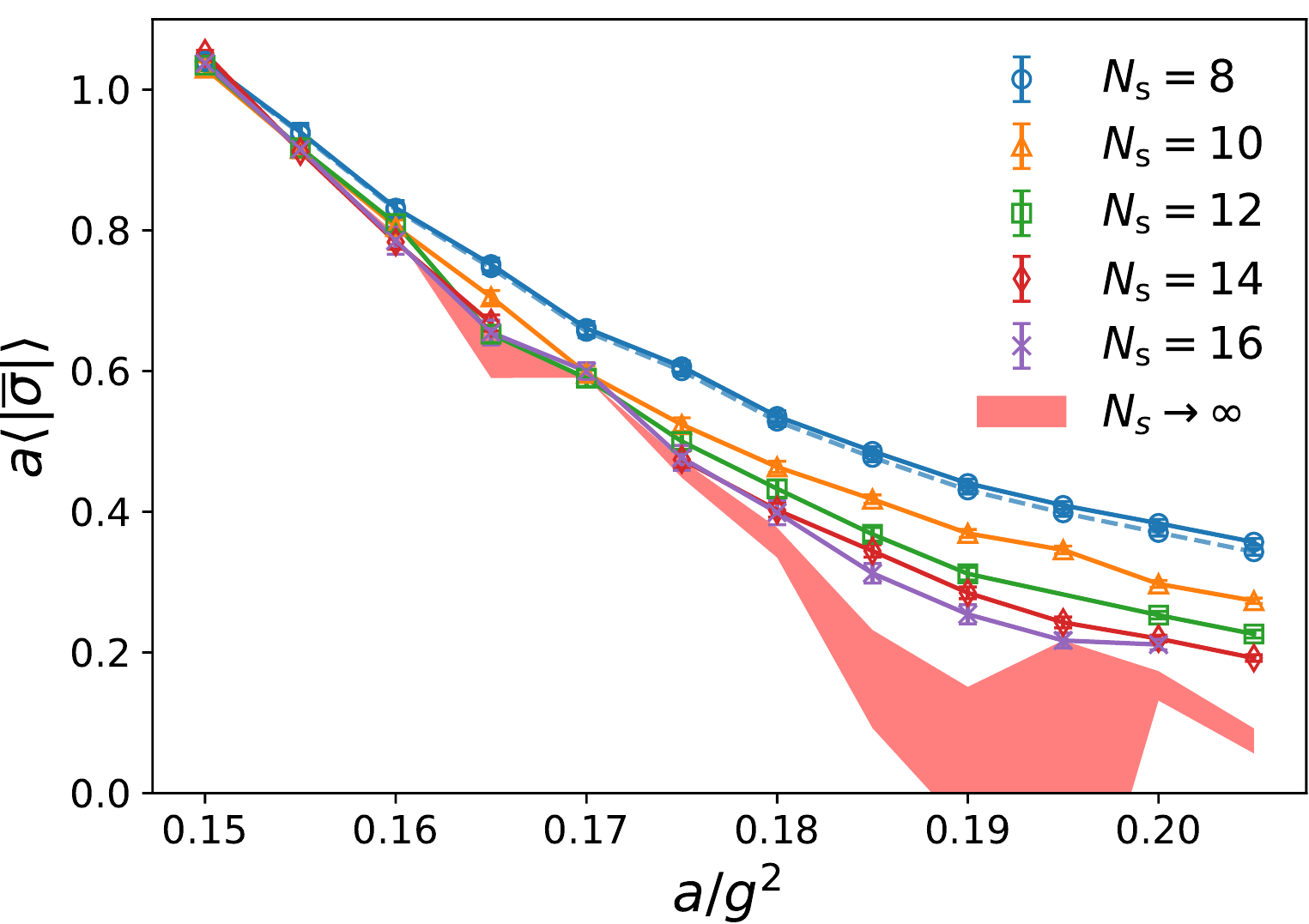}
	\caption{Coupling-dependence of the chiral condensate $\sig$ for various cubic lattice sizes, 
	$\Nt=\Ns$. The dashed line shows $\frac{-ag^2}{\Nf}\cc_{\mathrm{ov}}$ , as defined in \eqref{eq:chiral_condensate} (with the absolute value taken appropriately), for $\Ns=8$ and the red band is an extrapolation to the infinite volume. All quantities are 
	given in lattice units.}
	\label{fig:scale_setting}
\end{figure}

In this section we report our results obtained in the GN model in $2+1$ dimensions, using overlap
fermions for one reducible fermionic flavor, $\Nf=1$.

\subsection{Consistency checks}
As an important starting point, we test our discretization (\ref{eq:overlap}) and perform consistency 
checks with results in the existing literature. 
To this end, we show in \fref{fig:scale_setting} the dependence of 
the order parameter (\ref{eq:order_parameter}) on the coupling constant $g^2$ for increasing lattice 
volumes. The dashed blue line shows, exemplarily for the smallest lattice considered, the right-hand side of the Dyson-Schwinger equation (\ref{eq:chiral_condensate}) for comparison. This indicates that \eqref{eq:chiral_condensate} is, indeed, fulfilled. The coupling strengths we use for the bulk
of this work lie in the left half of \fref{fig:scale_setting}.

In \fref{fig:scale_setting} we also show an extrapolation to the infinite volume, using the finite-size scaling law
\begin{align}
	\sig = \alpha + \gamma L^{-\kappa}\;,
\end{align}
where $\alpha$, $\gamma$ and $\kappa$ are constants, for the $L$-dependence of the order parameter 
for every value of the coupling.

When $a/g^2$ takes values between $0.188$ and $0.198$ we find the offset 
$\alpha$ to be consistent with zero within errors in the infinite-volume limit,
which indicates the presence of a phase transition. In this case, $\kappa$ is related to the critical 
exponents $\beta$ and $\nu$ of the order parameter and correlation length, respectively, via
\begin{align}
	\kappa = \frac{\beta}{\nu}\;.
\end{align}
With this crude and naive method, we find $\beta/\nu=0.93\pm0.29$ as a weighted
average which -- while not competitive in precision -- is in quantitative
agreement with pertinent results obtained with dedicated methods as, for
example, collated in \cite{Sch17}, $\beta/\nu=0.62\dots0.85$. Recovering this
non-perturbative result is a strong indication that we are simulating the
correct physics. From now on we shall always consider strong enough couplings such that
chiral symmetry is spontaneously broken at $T\approx0=B$, i.e., we work in the strong-coupling or 
super-critical regime as in \sref{sec:analytical}.

\subsection{Vanishing magnetic field}
Having established the correctness of our method, we now present results for the
order parameter at vanishing magnetic field and non-zero temperature, which
allows for a comparison with results in \cite{HKK93_1,HKK93,KS01}, at least on a
qualitative level.

\begin{figure}[t]
	\includegraphics[width=0.95\linewidth]{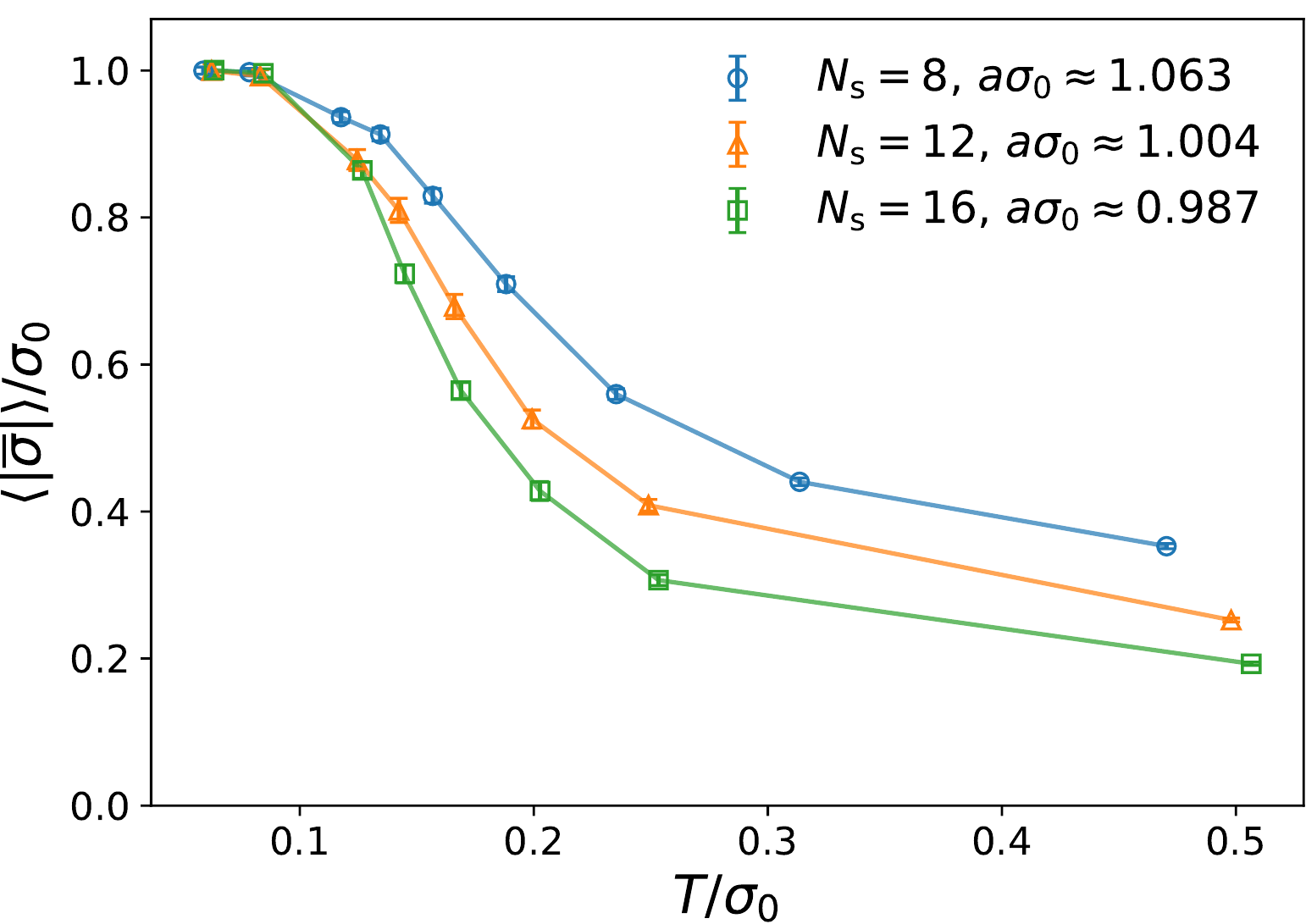}
	\caption{Temperature-dependence of the chiral condensate $\sig$ for different physical volumes and 
	$B=0$.}
	\label{fig:cc_vs_T}
\end{figure}

In \fref{fig:cc_vs_T} we show the $T$-dependence of $\sig$
for increasing physical volumes. We observe the 
expected spontaneous breaking of chiral symmetry at low temperatures, indicated by a non-vanishing 
order parameter, and a decrease of the condensate with increasing temperature, corresponding to the well-known 
picture of thermal fluctuations destroying long-range order and restoring chiral symmetry. Of course, 
as was mentioned above, $\sig$ cannot vanish exactly on finite volumes. What one can see, 
however, is that the phase transition becomes more pronounced as the volume increases, 
while the non-vanishing tail for high temperatures approaches lower and lower values. 

\begin{figure}[t]
	\includegraphics[width=0.95\linewidth]{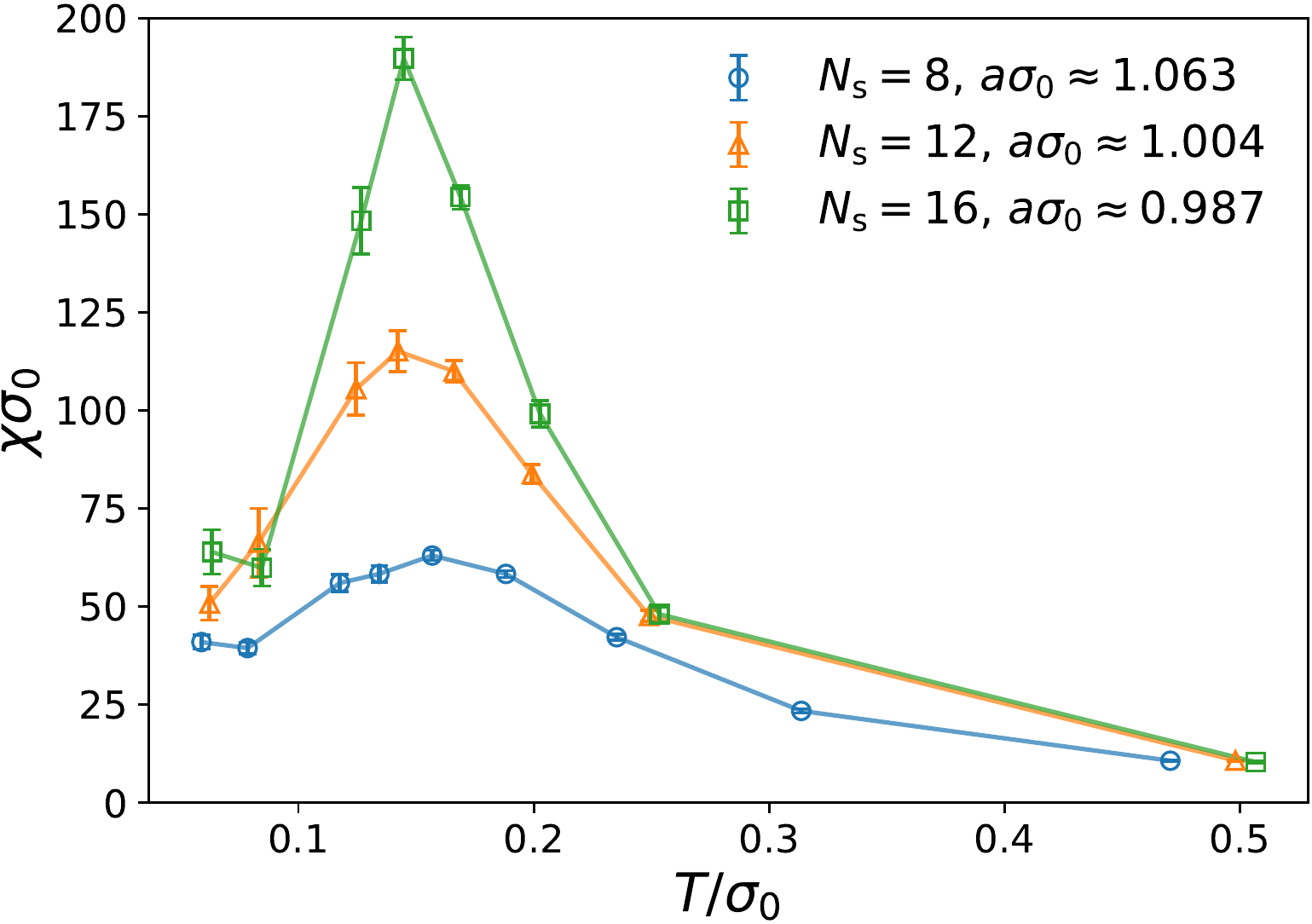}
	\caption{Chiral susceptibility (\ref{eq:susceptibility}) as a function of temperature for 
	different physical volumes.}
	\label{fig:chi_vs_T}
\end{figure}

In order to locate the phase transition we show in \fref{fig:chi_vs_T} the $T$-dependence 
of the chiral susceptibility (\ref{eq:susceptibility}) for different volumes. As expected, there is 
a pronounced peak at a critical temperature $T=\Tc$, which shifts slightly to lower temperatures as 
the volume is increased. For large enough volumes, where the peak becomes even more pronounced, we 
find $\Tc/\sigma_0\approx0.145$. 

We furthermore compute the Binder cumulant \cite{Bin81}, 
\begin{align}\label{eq:binder_cumulant}
	U_L := 1-\langle\bar{\sigma}^4\rangle/3\langle\bar{\sigma}^2\rangle^2\;,
\end{align}
as 
a function of $T$. The intersection of $U_L(T)$ for different volumes provides
us with another estimate for the critical temperature,
$T_c/\sigma_0\approx0.135$. We take the interval between the two values as a
rough estimate of the actual critical temperature. A direct comparison to the existing
literature \cite{HKK93_1,HKK93,KS01} is, unfortunately, not straightforward, as
those works either employ higher flavor numbers or use different scale settings. 

The observations presented so far are consistent with the GN model approaching a second-order phase 
transition in $T$ in the infinite-volume limit, as one would expect based on the large\,-$\Nf$ 
analysis of \sref{sec:analytical} and as has been previously observed in \cite{HKK93_1,HKK93,KS01}. 

Obviously, bosonic quantum fluctuations leave their mark 
on the system for flavor numbers as low as $\Nf=1$, as can be seen by comparing the critical temperature quoted above with its large\,-$\Nf$ value, $T_c/\sigma_0=1/2\ln(2)
\approx0.72$, the latter being significantly larger. This means that the broken phase 
shrinks when one departs from the mean-field limit by decreasing $\Nf$, which is not at all surprising given the tendency 
of quantum fluctuations to destroy any sort of long-range order. This phenomenon has 
also been observed in the earlier studies \cite{HKK93_1,HKK93,KS01} and occurs in the 
$(1+1)$-dimensional model as well \cite{LPW20}. 

We remark that even the largest volume considered in this work is still comparatively small. 
Thus, one should not be tempted to draw quantitative conclusions about the precise location or the 
order of the chiral phase transition at $\Nf=1$. The qualitative behavior, however, which is what we 
are ultimately interested in at $B\neq0$, is as expected, which further builds up confidence in 
the chosen discretization.

\subsection{Non-zero magnetic field}
\subsubsection{Temperatures close to zero}
\begin{figure}[t]
	\subfloat[Continuum extrapolation.]{
		\includegraphics[width=0.95\linewidth]{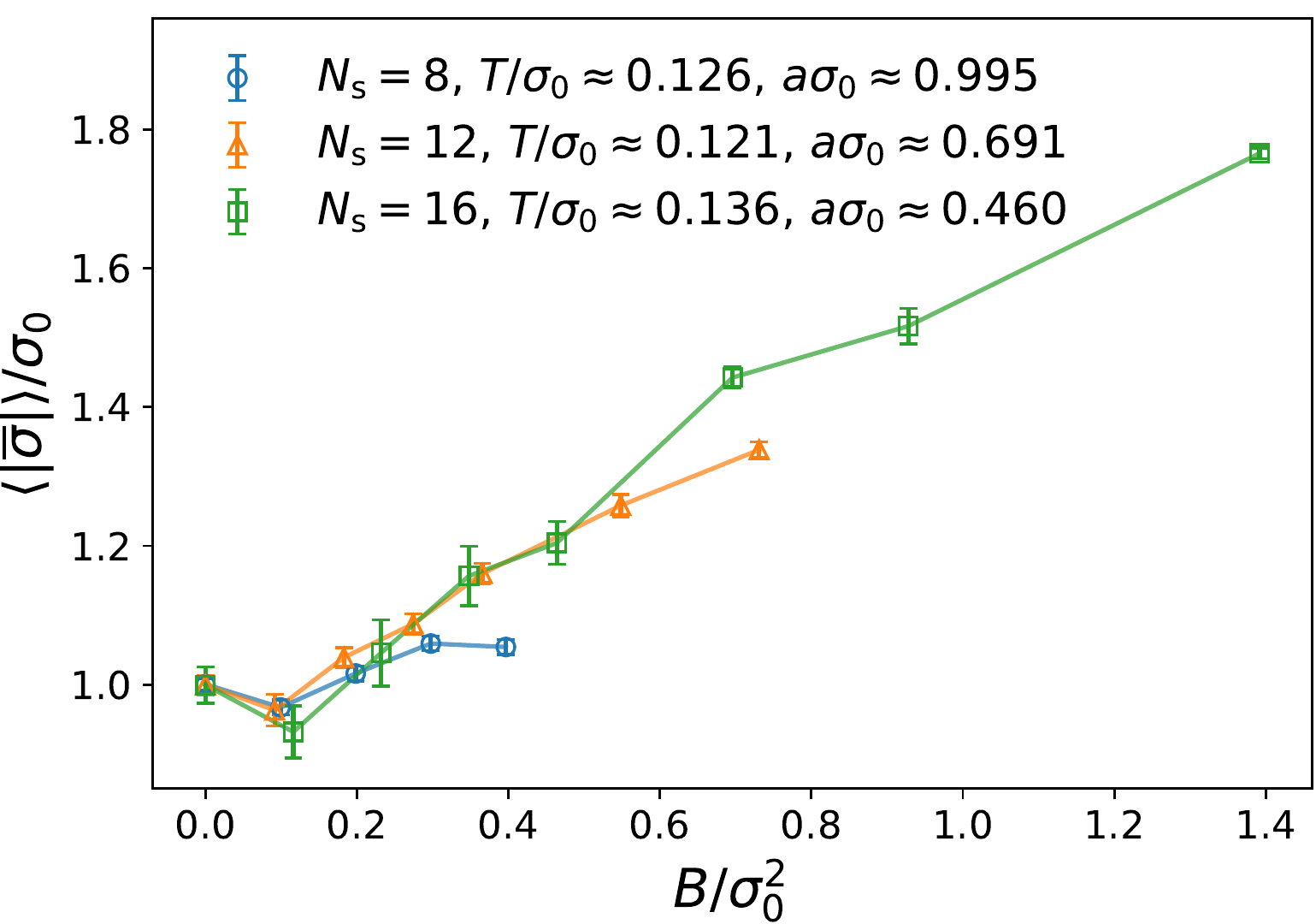}
		\label{fig:cc_vs_B_cont}}\\
	\hspace{-0.3cm}
	\subfloat[Infinite-volume extrapolation.]{
		\includegraphics[width=0.97\linewidth]{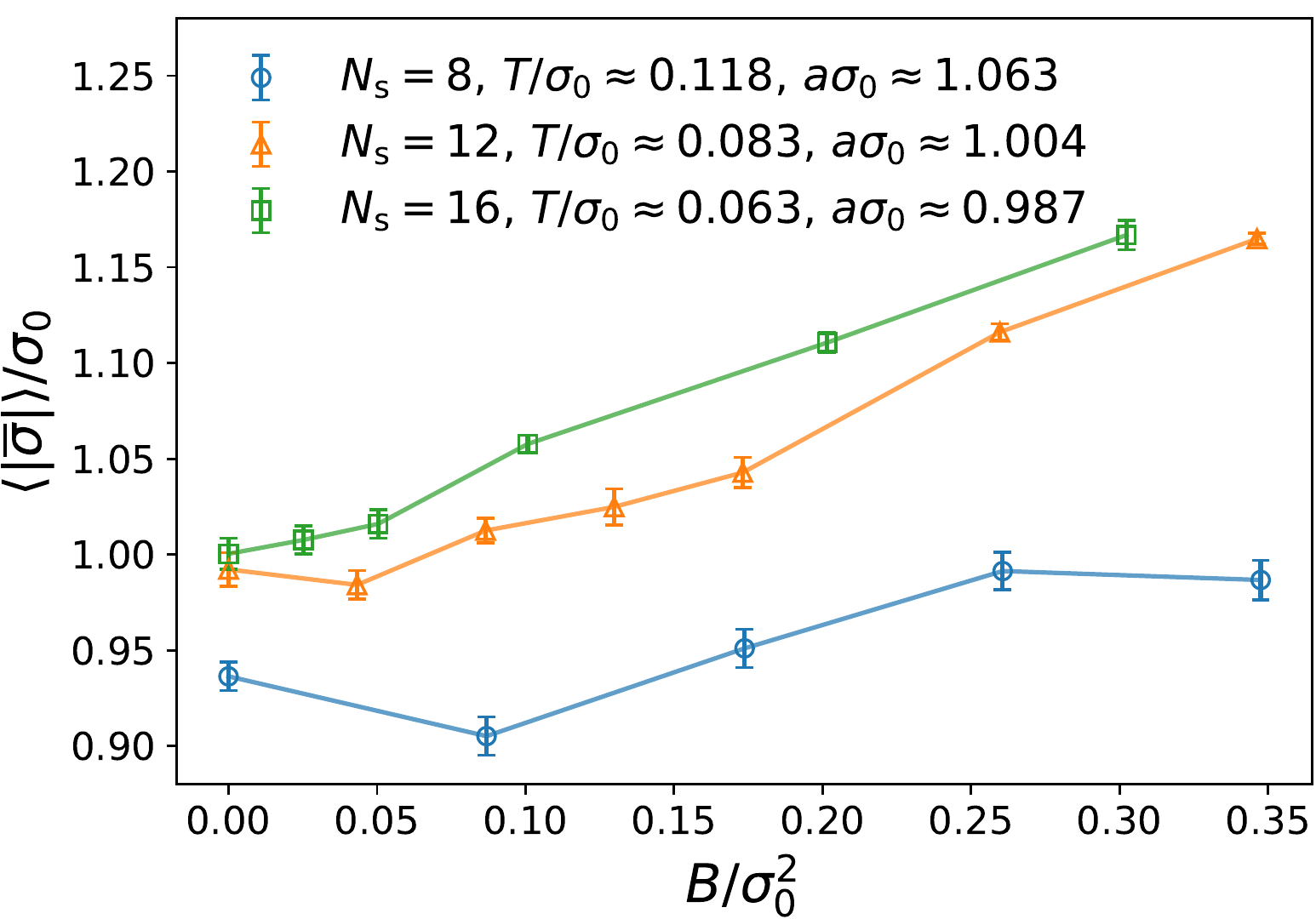}
		\label{fig:cc_vs_B_infVol}}
	\caption{Magnetic-field-dependence of the chiral condensate $\sig$ for low 
	temperatures.}
	\label{fig:cc_vs_B}
\end{figure}

We now switch on an external magnetic field and first devote our attention to the lowest available 
temperatures. The $B$-dependence of the chiral condensate for various different 
lattice constants and volumes is shown in Figs.~\ref{fig:cc_vs_B_cont} and \ref{fig:cc_vs_B_infVol}, 
respectively. In all data the magnetic field is found to increase the chiral
condensate which is in qualitative agreement with the large-$\Nf$ expectation.

While the latter predicts quadratic growth for our scenario (and only linear
growth in the sub-critical coupling regime), our data look rather linear but
might still be compatible with a weak quadratic growth. This discrepancy could
also come from discretization effects. Although \fref{fig:cc_vs_B_cont} suggests
that they may be small in the interacting theory, such a deviation
would be the expected form of discretization artifacts in the non-interacting
case as discussed in \aref{app:discretizations}. We find that such artifacts
would systematically diminish the chiral condensate such that we are confident
that our results are qualitatively correct even if discretization effects are
larger than suggested by \fref{fig:cc_vs_B_cont}.

Moreover, one observes curious non-monotonic behavior of 
$\sig$ with $B$, as the order parameter seems to assume a minimum at the 
lowest possible non-vanishing magnetic field, corresponding to $b=1$ in 
\eqref{eq:flux_quantization}, for all lattice spacings. For flux parameters larger than $1$ the 
condensate then grows monotonically with $B$.
This non-monotonicity, however, is a finite-size effect, as becomes clear by looking at the 
infinite-volume extrapolation shown in \fref{fig:cc_vs_B_infVol}, where $b=1$ ceases to be a minimum 
of $\sig$ for the largest available volume (green curve). We note that the physical volume 
considered in \fref{fig:cc_vs_B_cont}, which we keep approximately constant as we decrease the lattice 
spacing, corresponds to the smallest volume in \fref{fig:cc_vs_B_infVol}. 

\begin{figure*}[t]
	\subfloat{
		\includegraphics[width=0.3235\linewidth]{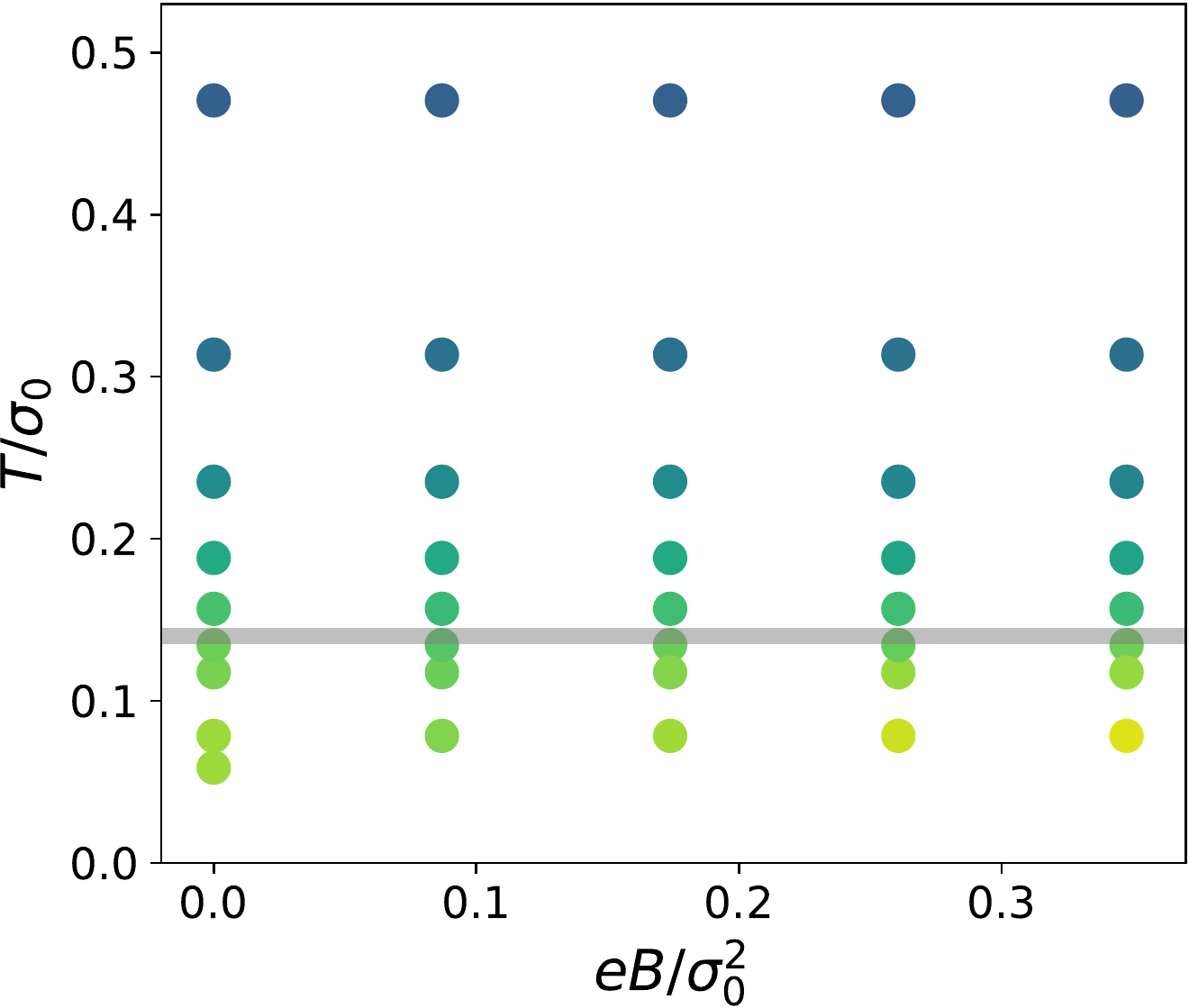}}\hspace{0.1cm}
	\subfloat{
		\includegraphics[width=0.2805\linewidth]{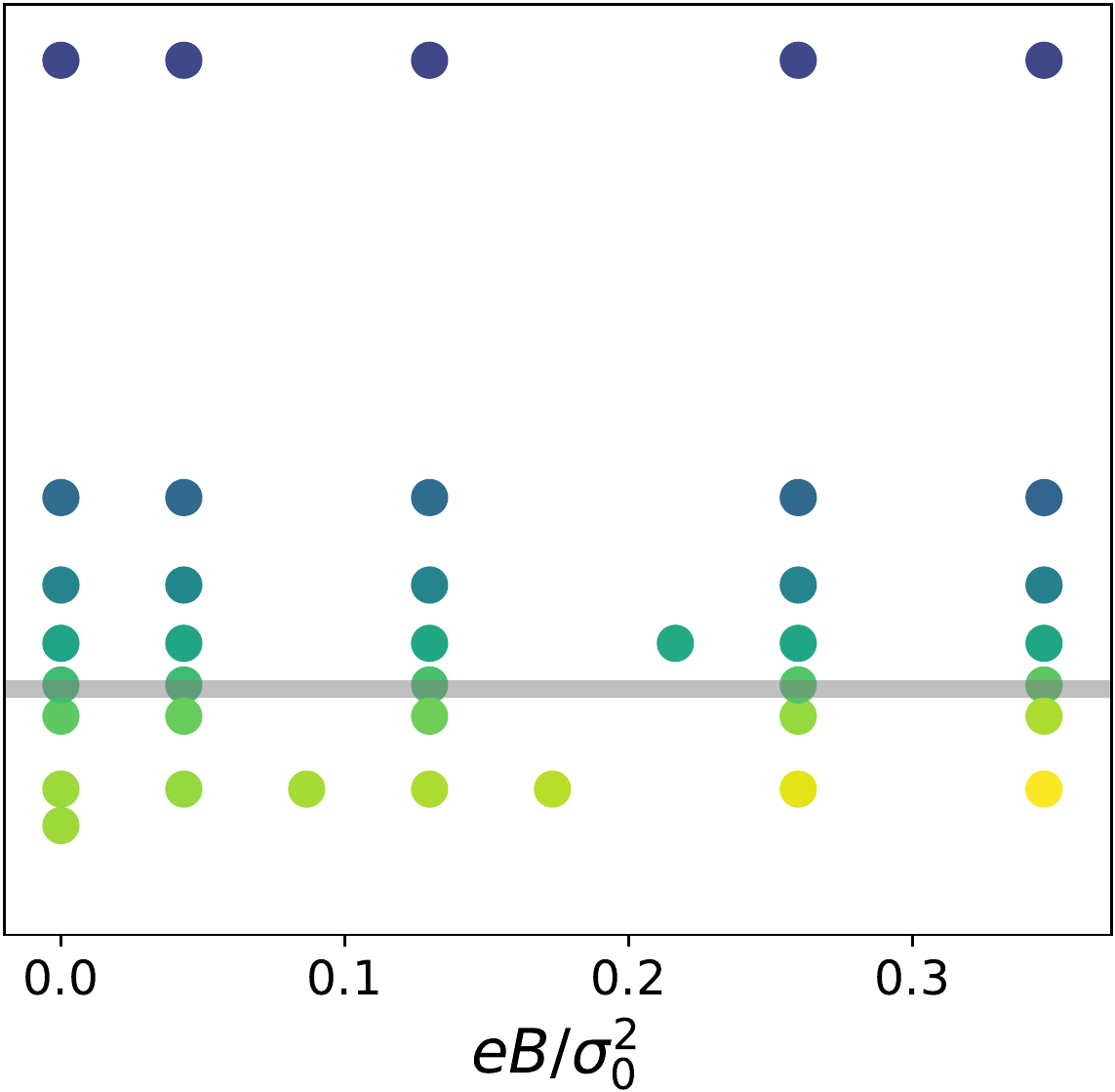}}\hspace{0.1cm}
	\subfloat{
		\includegraphics[width=0.344\linewidth]{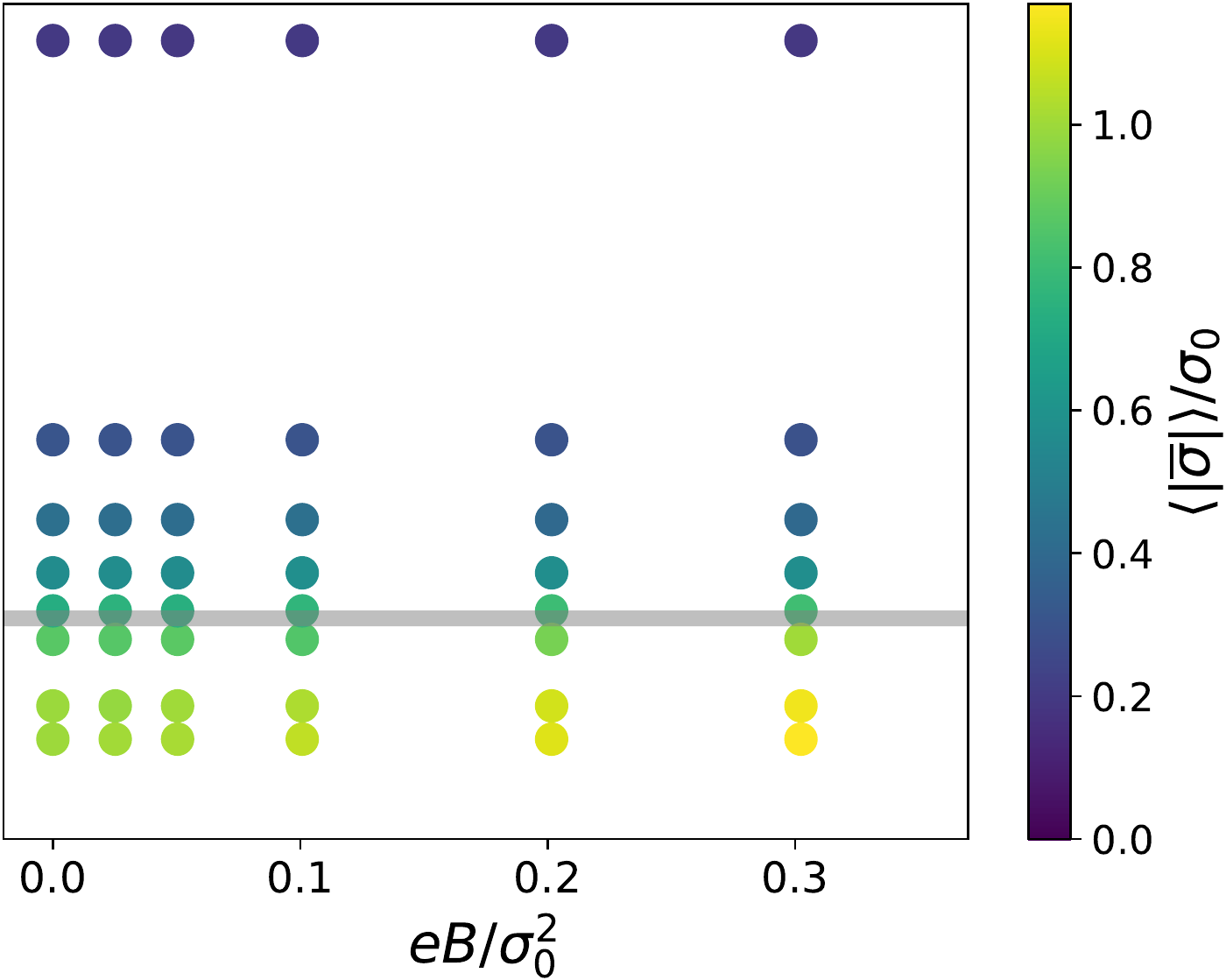}}
	\caption{$(B, T)$ phase diagrams for increasing volumes at constant lattice 
	spacing. Left: $\Ns=8$, $a\sigma_0\approx1.063$. Center: $\Ns=12$, $a\sigma_0\approx1.004$. Right: $\Ns=16$, $a\sigma_0\approx0.987$. The gray band indicates our estimate for the critical temperature at $B=0$ and on the 
	largest lattice, see the main text.}
	\label{fig:pd_infVol}
\end{figure*}

The largest magnetic fields we plot in \fref{fig:cc_vs_B} are determined by our requirement that 
$b\leq\Ns^2/16$. For larger magnetic fields we find unphysical saturation effects, the onset of which 
is already visible in the $\Ns=8$ data of \fref{fig:cc_vs_B} (blue curves). We plan to present a more 
detailed discussion of these discretization artifacts and the aforementioned finite-size effects, as 
well as a thorough spectral analysis of the overlap operator for the GN model in non-zero magnetic 
fields in a forthcoming publication.

We arrive at the conclusion that, on sufficiently large volumes and for temperatures close 
to zero, the magnetic field causes the order parameter to increase, thus enhancing the breaking 
of chiral symmetry, in accordance with the mean-field prediction of magnetic catalysis outlined 
in \sref{sec:analytical}. This is hardly surprising, given the effective one-dimensional dynamics 
induced by the magnetic field. In fact, as has been argued in \cite{GMS94}, magnetic catalysis at zero 
temperature is a universal, i.e., model-independent feature in $2+1$ dimensions, at least in the 
absence of gauge degrees of freedom \cite{MS02}.

\subsubsection{Higher temperatures}
Next, we study the combined influence of finite temperature and magnetic field on the order 
parameter. We show phase diagrams in the $(B, T)$ plane for various lattice sizes in 
\fref{fig:pd_infVol}.

Evidently, magnetic catalysis takes place not only for the lowest temperatures, but for all $T$ 
below $\Tc$. We indicate the values of $\Tc$ at $B=0$ and $\Ns=16$, determined above via $\chi$ and $U_L$, respectively, by the gray bands. For higher temperatures the magnetic 
field ceases to have a noticeable effect on the order parameter. This is not unexpected, as in this 
region we only measure the (modulus of the) fluctuations of $\sigma$ around zero due to our 
definition of $\sig$ in \eqref{eq:order_parameter}. 

\begin{figure}[b]
	\includegraphics[width=\linewidth]{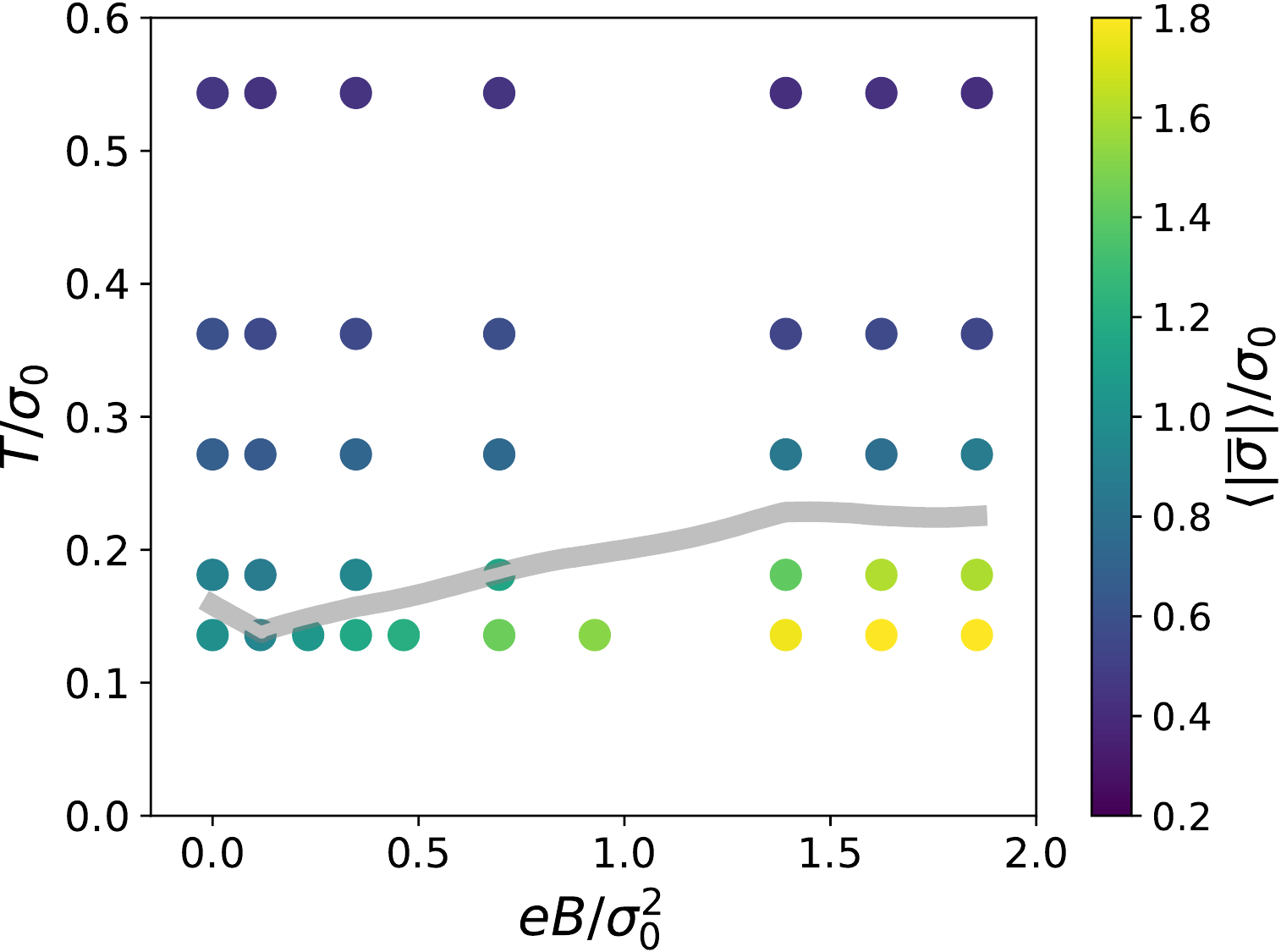}
	\caption{$(B,T)$ phase diagram for $\Ns=16$ and $a\sigma_0\approx0.460$. The gray band shows our crude estimate for the $B$-dependence of the critical temperature of the phase transition, see the main text. The scale on the color bar is different from \fref{fig:pd_infVol}.}
	\label{fig:pd_cont}
\end{figure}

The magnetic fields we restrict ourselves to (in order to avoid discretization effects) in our 
lattice simulations at fixed lattice 
spacing are 
quite small, $eB/\sigma_0^2\lesssim0.35$. Hence, the results obtained in the large\,-$\Nf$ 
approximation, shown in 
\fref{fig:pd_large_N}, suggest that one should not expect the broken region to grow in size all 
that much. Indeed, this expectation is confirmed by \fref{fig:pd_infVol}. 

To investigate larger values of $eB/\sigma_0^2$, we consider the $(B,T)$ phase diagram for the 
smallest available lattice spacing in \fref{fig:pd_cont}. One observes that for strong enough 
magnetic fields the region of spontaneously broken chiral symmetry indeed starts to grow, as expected 
from \fref{fig:pd_large_N}. We roughly indicate this by the gray band, which shows the critical 
temperature $\Tc$, determined by the susceptibility (\ref{eq:susceptibility}), as a function of 
$B$. When $\Tc$ cannot be determined unanimously we take the average of the two temperatures 
corresponding to the competing peaks instead and we do not show error bars for the resulting -- very crude -- estimate. 
Recall that finite-volume effects distort the behavior for weak magnetic fields.

It would be interesting for future studies to consider even stronger 
magnetic fields in order to compare Figs.~\ref{fig:pd_large_N} and \ref{fig:pd_cont} on a more 
quantitative level. In conjunction with simulations at different flavor numbers, one could aim at 
finding a relation between the phase boundaries as $\Nf$ is varied. In the simplest scenario, the 
critical temperature $\Tc(B)$ could conceivably be related to its large\,-$\Nf$ value by a mere 
$\Nf$-dependent scaling factor.

\subsection{Search for inhomogeneous phases}

Finally, we investigate the existence of inhomogeneous phases by studying the spatial correlator 
(\ref{eq:spatial_correlator}). Such a phase would likely occur at low temperatures and 
relatively strong magnetic fields, the former since thermal fluctuations will wash out any 
inhomogeneities and the latter since we know that the order parameter is homogeneous for vanishing 
magnetic field \cite{BKW21,PWW22,WP22}. 

\begin{figure}[t]
	\includegraphics[width=0.95\linewidth]{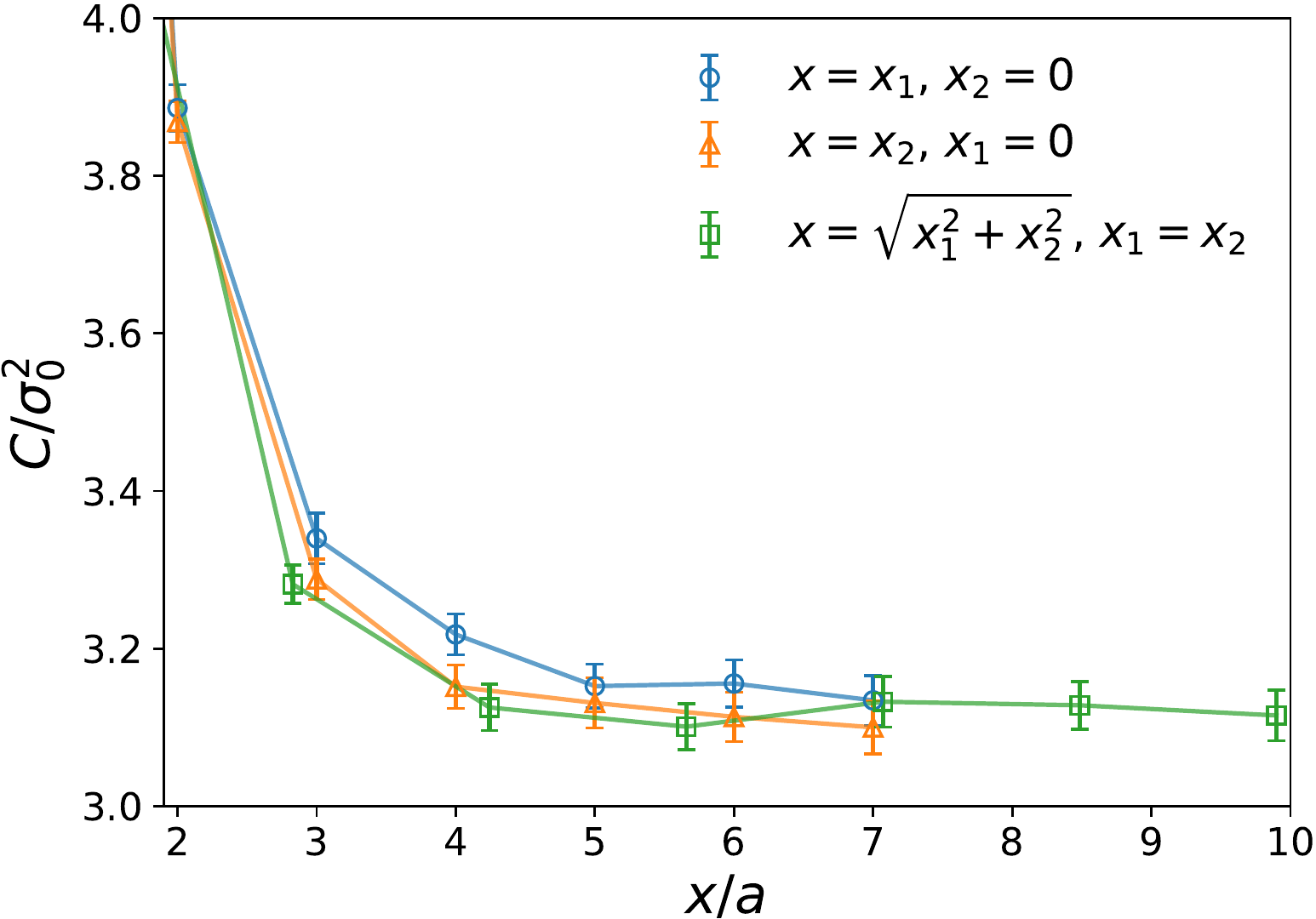}
	\caption{Spatial correlator (\ref{eq:spatial_correlator}) for $\Ns=\Nt=16$, $eB/\sigma_0^2\approx1.39$ and $a\sigma_0\approx0.460$ along the two coordinate axes and their diagonal. Due to the lattice periodicity we only show the first half of the respective abscissas.}
	\label{fig:spatial_correlator}
\end{figure}

\fref{fig:spatial_correlator} shows the correlator $C$ from \eqref{eq:spatial_correlator} 
for a strong magnetic field along the two spatial coordinate axes and their
diagonal. Each of them decays monotonically to a constant close to the
contribution from the disconnected terms. In fact, we
can showcase the rotational invariance here and no further structure is seen in
other directions or for other parameters. We conclude that the assumption of
spatial homogeneity is well justified in the accessible parameter range. Whether
stronger magnetic fields could induce a spatially varying order parameter,
especially in combination with a finite chemical potential, is a question for
future studies.

	\section{Discussion}\label{sec:discussion}
	We have investigated the $(2+1)$-dimensional Gross-Neveu model (\ref{eq:lagrangian}) exposed to an 
external 
magnetic field in the chiral limit using one reducible flavor of fermions and Neuberger's 
formulation (\ref{eq:overlap_massless}) of the overlap operator. The auxiliary scalar field 
$\sigma$ couples in a way so as to preserve the continuum Ward identity (\ref{eq:ward_identity}) 
relating the expectation value of $\sigma$ to the chiral condensate. 

Our results suggest that the magnetic catalysis phenomenon, i.e., an enhancement of the order 
parameter for chiral symmetry breaking with the magnetic field, persists for finite flavor numbers, 
in accordance with the phenomenological picture of the magnetic field reducing the number of spatial 
dimensions, thus promoting infrared dynamics. On small volumes, however, this effect is non-monotonic 
for weak magnetic fields. We also remark that our lattice formulation seems to suffer from strong
discretization effects in a free-theory setup, while the interacting case appears less problematic.

We have furthermore investigated the fate of magnetic catalysis at finite temperature and found 
that it persists for all temperatures below the phase transition. Our findings are thus in 
qualitative agreement with mean-field \cite{Kli91,Kli92,Kli92_2} as well as 
beyond-mean-field \cite{KPR13,SG12} calculations. The phase of spontaneously broken chiral 
symmetry grows slightly for the strongest magnetic fields considered but 
shrinks overall in comparison to the large-$\Nf$ limit.

It is important to stress that our results are very different from the well-known \emph{inverse 
magnetic catalysis} effect, i.e., a decrease of the order parameter with $B$, that takes place in 
QCD at temperatures close to the chiral crossover \cite{BBE12,BBE12_2}. In QCD, the 
critical temperature furthermore decreases with the magnetic field \cite{End15,DMN18}, which we  do not observe either. We now comment on this issue.

In QCD the aforementioned effects are likely caused by a delicate interplay between quark and 
gluonic degrees of freedom \cite{BEK13}, which our model, lacking the latter, cannot reproduce. Thus, one should not be tempted to interpret our results as new physics. Rather, we argue that the GN 
model is simply, and unsurprisingly, insufficient for a proper description of QCD once gluonic 
effects become important. While our results are thus in agreement with the expectation, we also 
stress that it was not entirely clear to us before this study to what extent the effect of quantum 
fluctuations might change the qualitative picture. 

As was mentioned in the Introduction, we believe that our 
work serves as a starting point for the ultimate goal of studying QCD in background magnetic fields 
from the point of view of effective models beyond the mean-field limit and on the lattice. In the 
following we discuss ways to systematically improve the GN model in order to approach QCD. 
To this end, one should first consider models in $3+1$ dimensions that have the same continuous chiral symmetry as QCD. One may then take 
gluonic interactions into account, for example, by coupling the fermions to the Polyakov loop 
\cite{MO96}. Most importantly, the crucial back-reaction of magnetized quarks onto the gluonic 
distribution can be taken into account by introducing a suitable effective $B$-dependent coupling. 
This has been shown to reproduce the desired features of QCD in \cite{FGK14,EM19,FCL14,TFA21}. 

One could furthermore consider endowing the scalar fields in the NJL model with kinetic terms, thus 
enabling their interpretation as dynamical mesons and potentially add quartic mesonic 
self-interactions as well. The ensuing linear sigma model coupled to quarks (LSMq) has the added 
advantage 
that it is renormalizable in $3+1$ dimensions, whereas the GN and NJL models are not. If 
one then incorporates the aforementioned magnetic-field-dependent couplings, while properly taking 
into account plasma screening effects, one also observes inverse magnetic catalysis as in QCD 
\cite{ALZ15,ADH15,AHL21}. 

A proper understanding of such effective theories for QCD beyond the mean-field limit, e.g., from 
\emph{ab initio} lattice simulations at finite numbers of quark flavors and colors is therefore 
certainly desirable. For reviews on the topic of reproducing features of QCD in magnetic fields using 
model theories and a more complete list of references, see \cite{AHL21,And21r,BF21}.

We briefly comment on possible implications for condensed-matter systems that are
described by four-Fermi theories. While in this work we are only concerned with the strong-coupling 
regime, in 
which chiral symmetry is broken at zero temperature and magnetic field, we believe that the 
qualitative predictions of mean-field studies should also remain valid for weak 
couplings. This would then imply that strong enough magnetic fields are indeed capable of generating a 
mass gap, providing further evidence \cite{SSW98,FGI03} that magnetic catalysis could be 
responsible for the kink-like behavior of the thermal conductivity of superconducting cuprates 
exposed to a magnetic field observed in \cite{KOG97}.

Finally, our results suggest that a small magnetic field does not seem to induce inhomogeneous phases in 
the GN model in $2+1$ dimensions at zero density. A detailed study of the finite-density case is 
currently underway.

Our simulation results as well as the tools required to reproduce the figures
shown in this work are available online \cite{data,code}.

	\begin{acknowledgments}
		We thank Björn Wellegehausen for providing the code base used in the present 
		work and for useful discussions regarding the implementation of the overlap 
		operator in our setup. M.M. thanks Georg Bergner, Gergely Endr\H{o}di, Tam\'as Kov\'acs 
		and Ivan Soler for enlightening discussions. J.J.L. thanks Ed Bennett for
    helpful discussions about the reproducibility and openness of this
    publication. 

    This work has been funded by 
		the Deutsche Forschungsgemeinschaft (DFG) under Grant No. 406116891 within 
		the Research Training Group RTG 2522/1. The simulations were performed on 
		resources of the Friedrich Schiller University in Jena supported in part by 
		the DFG Grants INST 275/334-1 FUGG and INST 275/363-1 FUGG. The work of J.
    J. L. was partly supported by the UKRI Science and Technology Facilities
    Council (STFC) Research Software Engineering Fellowship EP/V052489/1 and by
    the Supercomputing Wales project, which is part-funded by the European
    Regional Development Fund (ERDF) via Welsh Government. 

    This work would never have been possible without the great \verb|python|
    ecosystem for scientific computing \cite{python}. For our analyses, we
    explicitly imported the packages
    \cite{numpy,scipy,pandas,matplotlib,mpmath,statsmodel}
    but we are also grateful for creation and maintenance of all their
    dependencies.
  \end{acknowledgments}

  \section*{Open Access Statement}
  For the purpose of open access, the authors have applied a Creative Commons Attribution (CC BY) licence to any author accepted manuscript version arising.

  \section*{Data Availability Statement}
  Full data underlying this work are available in Ref.~\cite{data}. Fully
  automated analysis workflows are available in Ref.~\cite{code}. Raw data and
  the simulation code for generating the configurations are available upon
  request.
	
	\appendix
	\section{Derivation of the effective potential in the large\texorpdfstring{\,-$\Nf$}{-Nf} 
	limit}\label{app:large_N}
	In this appendix we outline the calculation of the effective potential in 
\eqref{eq:effective_potential}, see also \cite{BVW91}. The main difficulty
is, of course, the fermionic determinant, $\det(D)$. For the derivation
we shall, in fact, consider the more general Dirac operator
\begin{align}
	D = \slashed{\partial} + \ii e\slashed{A} + \sigma + \mu\gamma_0\;,
\end{align}
where we have also included a chemical potential $\mu$ and $A_\mu$ is given in \eqref{eq:continuum_vector_potential}. In the following we assume $B>0$ without loss 
of generality. 

For the computation of $\ln\det D$ we use the zeta-function regularization method \cite{Haw77}:
\begin{align}\label{eq:log_det_zeta}
	\ln\det D = \frac{1}{2}\ln\det D^2 = -\frac{1}{2}\frac{\partial}{\partial s}\zeta_{D^2}(s)
	\bigg\vert_{s=0}\;,
\end{align}
with the zeta-function of $D^2$ defined by
\begin{align}\label{eq:zeta_function}
  \zeta_{D^2}(s) = \frac{1}{\Gamma(s)}\int_0^\infty \!\mathrm{d}t\, t^{s-1} \tr{e^{-t D^2}}\;,
\end{align}
where $\Gamma(s)$ denotes the usual gamma function. The spectrum of $D^2$ is known and its eigenvalues read
\begin{align}\label{eq:eigenvalues_finite_B}
	\lambda = \sigma^2 + (\omega_n+\ii\mu)^2 + (2l + 1 + \alpha)eB\;,
\end{align}
where $\omega_n = \frac{\pi}{\beta}(2n+1)$ are the Matsubara frequencies ($n\in\Z$), 
$l\in \mathbb{N}_0$ 
denotes the Landau level index, and $\alpha\pm1$ denotes the Zeeman splitting of energy levels of 
fermions with opposite spin due to the Pauli term in $D^2$. The eigenvalues come with a degeneracy of 
$2\cdot\frac{VeB}{2\pi\beta}$, where the first factor of $2$ comes from the use of a reducible 
representation of gamma matrices while the second factor is the standard Landau level 
degeneracy. 

We are thus left with
\begin{align}
	\begin{aligned}
		\frac{\zeta_{D^2}(s)}{V} =
    \frac{1}{\Gamma(s)}\frac{eB}{\pi\beta}\bigg[\int_0^\infty\! 
    \mathrm{d}t\,t^{s-1}	e^{-t\sigma^2}\sum_{n=-\infty}^\infty e^{-t(\omega_n+\ii\mu)^2}+\\
    2\int_0^\infty\! \mathrm{d}t\,t^{s-1}e^{-t\sigma^2}\sum_{n=-\infty}^\infty e^{-t(\omega_n+\ii\mu)^2}
		\sum_{l=1}^{\infty}e^{-2eBlt}\bigg]\;,
	\end{aligned}
\end{align}
where we have already performed the sum over $\alpha$ and split up the summation over Landau 
levels into magnetic-field-independent terms ($l=0$) and corrections due to $B$ ($l>0$). By 
performing a Poisson resummation in $n$ and taking the integrals over $t$, a 
straightforward calculation leads to an expression for the zeta function, whose derivative with 
respect to $s$ at $s=0$ simplifies to

\begin{align}\label{eq:zeta_function_derivative}
	\begin{aligned}
		\frac{1}{V}\frac{\partial}{\partial s}\zeta_{D^2}(s)&\bigg\vert_{s=0} = \frac{eB}{\pi}
		\vert\sigma\vert - 
		\frac{(2eB)^{3/2}}{\pi}\zeta_H\left(-\frac{1}{2}, \frac{\sigma^2}{2eB}\right)\\
		-\frac{eB}{\pi\beta}\sum_{l=0}^\infty d_l\bigg[&\ln\left(1+e^{-\beta\left(
		\sqrt{\sigma^2+2eBl}+\mu\right)}\right) + (\mu\leftrightarrow-\mu)\bigg]\;,
	\end{aligned}
\end{align}
where $d_l=2-\delta_{l0}$. After setting $\mu=0$ and inserting this expression into 
(\ref{eq:log_det_zeta}) and (\ref{eq:effective_potential}), we obtain
\begin{align}
\notag	\frac{V_\mathrm{eff}}{V} 
		= &\frac{\sigma^2}{2g_R^2} 
		 - \frac{\sqrt{2}}{\pi}(eB)^{3/2}\zeta_H\left(-\frac{1}{2}, \frac{\sigma^2}{2eB}\right)
	     + \frac{\vert\sigma\vert eB}{2\pi} \\
	    - &\frac{eB}{\pi\beta}\sum_{l=0}^{\infty}d_l\ln\left(1+\exp\left(-\beta\sqrt{\sigma^2+2eBl}
	    \right)\right)\;,
\end{align}
where we have replaced $g^2$ by the renormalized coupling $g_R^2$ as dictated by the zeta function 
formalism. Finally, we introduce the minimum of the effective potential at vanishing 
temperature, density and magnetic field, $\sigma_0=-\pi/g_R^2$, to recover 
(\ref{eq:veff_closed_form}).

	\section{Comparison of fermion discretizations}\label{app:discretizations}
	We discuss and compare three different fermion discretizations one could employ when 
attempting to study 
the GN model exposed to magnetic fields: naive, SLAC and 
overlap fermions. For the comparison we consider a theory of massive 
non-interacting fermions in an 
external magnetic field, characterized by the Lagrangian
\begin{align}
	\L = \psibar\left(\slashed{\partial}+\ii e\slashed{A} + m\right)\psi =: 
	\psibar D\psi\;,
\end{align}
and compute the chiral condensate
\begin{align}
	\cc := -\frac{1}{V}\frac{\partial}{\partial m}\ln Z\;,
\end{align}
where the partition function $Z$ is given by the fermion determinant,
\begin{align}
	Z  = \det D\;.
\end{align}

We have already computed $\ln\det D$ in the continuum theory in 
\aref{app:large_N}, allowing us to 
directly use the result (\ref{eq:zeta_function_derivative}) by setting 
$\sigma=m>0$. Thus, 
with \eqref{eq:log_det_zeta}, the chiral condensate in the continuum at $
\mu=0$ is given by the closed-form expression
\begin{align}\label{eq:condensate_free_theory_continuum}
	\begin{aligned}
		\cc(B) &= \\
		\frac{eB}{2\pi} - 
		\frac{m}{\pi}&\sqrt{\frac{eB}{2}}
		\zeta_H\left(\frac{1}{2}, \frac{m^2}{2eB}\right) + 
		\frac{eB}{\pi}\sum_{l=0}^{\infty}
			\frac{d(l)m}{\varepsilon_l}\frac{1}{1+e^{\beta\varepsilon_l}}\;,
	\end{aligned}
\end{align}
where
\begin{align}
	\varepsilon_l = \sqrt{\displaystyle \vphantom{\mathbf{p}^2}m^2+2eBl}\;.
\end{align}
We remark again that the volume-dependence only enters via the discretization of 
$eB$, \eqref{eq:flux_quantization}. This means that if one were to naively 
take the limit $eB\rightarrow0$ in a continuous manner, one would simultaneously 
approach the infinite-volume limit. 

To obtain the chiral condensate for vanishing magnetic field on a finite volume,
one must repeat the calculation leading up to 
\eqref{eq:condensate_free_theory_continuum}, replacing the last term in 
\eqref{eq:eigenvalues_finite_B} by $\mathbf{p}^2=p_1^2+p_2^2$, with 
$p_i=\frac{2\pi}{L}n_i$ and $n_i\in\Z$, and taking the sum over momenta in the place
of 
Landau levels. Taking the four-fold degeneracy of the eigenvalues into account and 
repeating the steps outlined in \aref{app:large_N} leads to the expression
\begin{align}\label{eq:condensate_free_theory_continuum_vanishing_B}
	\begin{aligned}
		\cc(B=0) &= \\
			\frac{m^2}{\pi} - 
			\frac{2m}{L^2}&\sum_{\mathbf{p}\neq\mathbf{0}}
			\frac{1}{\vert \mathbf{p}\vert}
			e^{-\frac{L^2m}{2\pi}\vert \mathbf{p}\vert} 
			+ \frac{4}{L^2}\sum_{\mathbf{p}}\frac{m}{\varepsilon_{\mathbf{p}}}
			\frac{1}
			{1+e^{\beta\varepsilon_{\mathbf{p}}}}\;,
		\end{aligned}
\end{align}
with
\begin{align}
	\varepsilon_{\mathbf{p}} = \sqrt{\displaystyle m^2+\mathbf{p}^2}\;,
\end{align}
for the chiral condensate on a finite volume and for $B=0$. Let us now turn to the lattice computations.

The basic ingredients for implementing external magnetic fields on the lattice are outlined in 
\sref{sec:lattice_magnetic_fields}. For naive and overlap fermions we use the formalism developed 
there, mainly involving the $U(1)$ gauge links in \eqref{eq:gauge_links_magnetic_field}, which enter 
the naive Dirac operator,
\begin{align}\label{eq:naive_operator}
	\Dnaive = \frac{1}{2}\gamma_\mu\left(\nabla_\mu^*+\nabla_\mu\right) + m\;, 
\end{align}
directly ($\nabla_\mu$ and $\nabla_\mu^*$ are defined in \eqref{eq:covariant_difference}) and the 
overlap operator via its Wilson kernel (\ref{eq:wilson_kernel}). 

When using the SLAC derivative, however, one cannot use compact gauge variables in the form of 
group-valued lattice links connecting neighboring lattice sites because the derivative itself is 
non-local and thus involves all lattice points in a given direction. We therefore briefly discuss an 
alternative solution: In analogy to the continuum, we define the SLAC Dirac operator as
\begin{align}\label{eq:slac_magnetic_field}
	\Dslac := \slashed{\partial}^\mathrm{SLAC} + \ii e\slashed{A} + m\;,
\end{align}
where the SLAC derivative in position space is given by the Toeplitz matrix \cite{KLW05}
\begin{align}
	\partial^\mathrm{SLAC}_\mu(x,y) = (-1)^{(x_\mu-y_\mu)/a}\frac{\pi/L_\mu}{\sin\left(\pi(x_\mu-
	y_\mu)/L_\mu\right)}
\end{align}
if $x_\mu\neq y_\mu$ and $x_\nu=y_\nu$ for all $\nu\neq\mu$, and $\partial^\mathrm{SLAC}_\mu=0$ 
otherwise. 

Obviously, the discretization (\ref{eq:slac_magnetic_field}) is not gauge invariant. One could, 
however, attempt to treat the $e\slashed{A}$ term as a small perturbation if the magnetic field is 
not too large, such that (\ref{eq:slac_magnetic_field}) still describes the correct physics 
approximately.\footnote{One should note that this assumption is hard to justify
given that the gauge field is a linear function of $x$ that (at individual sites) can
have a magnitude proportional to $L$.} We do so in the following, reducing its numerical value as much as possible by 
employing the symmetric gauge
\begin{align}\label{eq:symmetric_gauge}
	A_0(x) = 0\;, \quad A_1(x) = -\frac{B}{2}x_2\;, \quad A_2(x) = \frac{B}{2}x_1\;,
\end{align}
with $x_{1,2}$ in the range $\big[\frac{-L}{2},\frac{L}{2}\big)$. Problems will inevitably 
arise once the kinetic momentum $p_\mu + eA_\mu$ crosses the boundary of the first Brillouin zone 
since there the SLAC derivative is discontinuous. This is the reason SLAC fermions are not used in 
gauge theories, and in our case such a crossing will occur for strong magnetic fields. 

We note that the minimal coupling prescription used in (\ref{eq:slac_magnetic_field}) makes the 
lattice boundary correction terms introduced in \eqref{eq:lattice_vector_potential} obsolete, as they 
cannot be compensated for in the absence of compact periodic gauge variables. We have verified that 
their inclusion indeed gives worse results. Notice, however, that we are dealing 
with a different physical situation with SLAC fermions as now the total magnetic flux through the lattice vanishes, see \sref{sec:lattice_magnetic_fields}.

Let us now compare the continuum result (\ref{eq:condensate_free_theory_continuum}) with the lattice 
chiral condensate, defined by
\begin{align}\label{eq:chiral_condensate_lattice}
	\cc_\mathrm{latt} = -\frac{1}{V}\tr\left[D_\mathrm{latt}^{-1}\right]\;,
\end{align}
where $\Dlatt$ stands for
\begin{align}
	\Dlatt = \begin{cases}\Dnaive  &\textnormal{for naive fermions}\;,\\
					 \Dslac  &\textnormal{for SLAC fermions}\;,\\
					 D\left(\id-\frac{a}{2}\Dov\right)^{-1} &\textnormal{for overlap fermions}\;,
	    \end{cases}
\end{align}
the operators $\Dov$ and $D$ being defined in Eqs. (\ref{eq:overlap_massless}) and 
(\ref{eq:overlap}), respectively. For naive fermions (\ref{eq:chiral_condensate_lattice}) has to be 
divided by the number of doublers, i.e., $8$ in $2+1$ dimensions, in order to compare with continuum 
results.

\begin{figure}[h]
	\includegraphics[width=\linewidth]{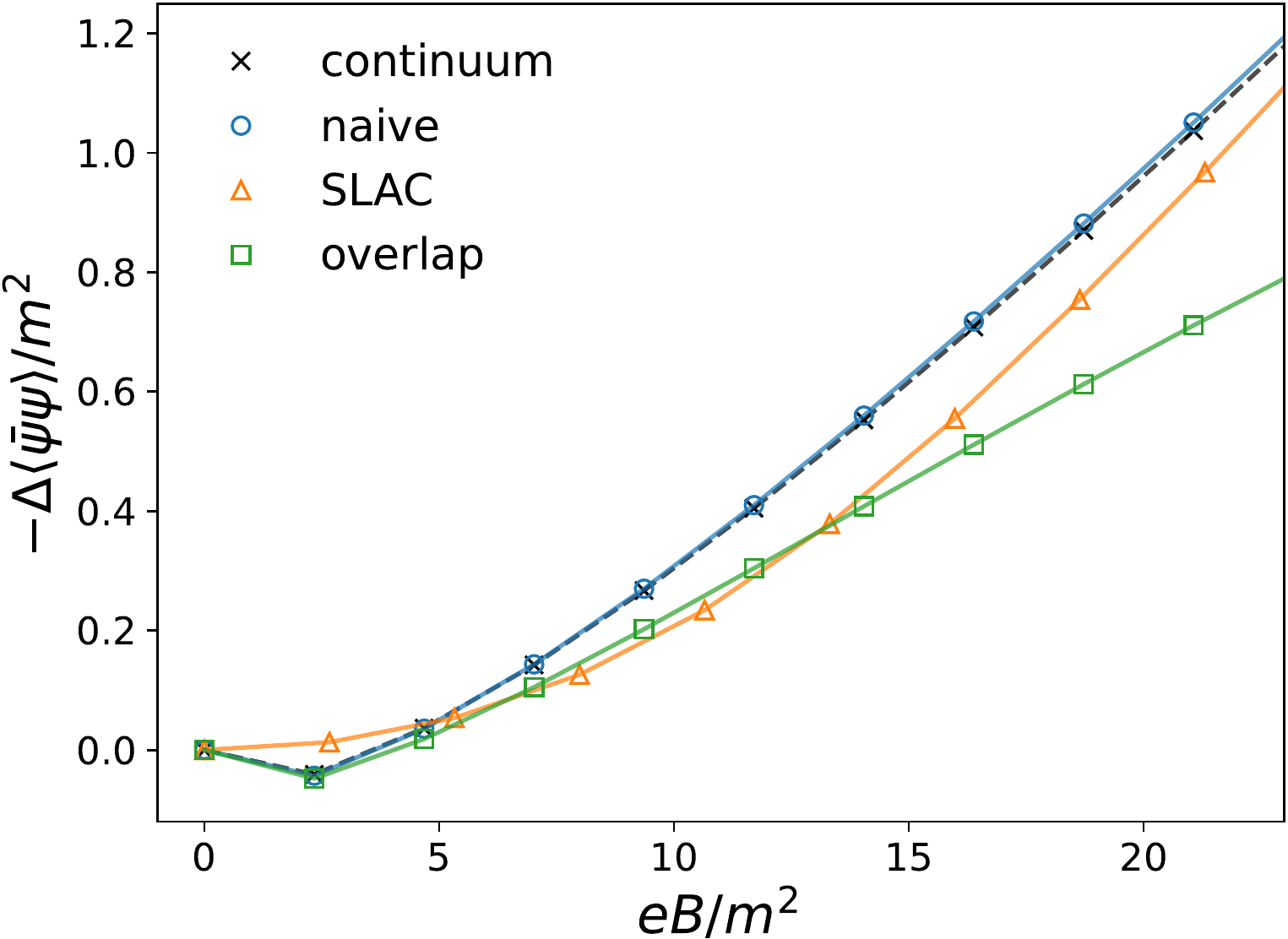}
	\caption{Comparison of $\Delta\cc$ in \eqref{eq:chiral_condensate_difference} between continuum 
	and lattice results for $\Ns=16$ (for the SLAC result we use $\Ns=15$), $\Nt=16$ and $am=0.1024$. 
	Notice that, since we work in a finite volume, the magnetic field is discrete even in the 
	continuum.}
	\label{fig:condensate_free}
\end{figure}

We show in \fref{fig:condensate_free} the change in the chiral condensate induced by the magnetic 
field,
\begin{align}\label{eq:chiral_condensate_difference}
	\Delta\cc = \cc(B) - \cc(0)\;,
\end{align}
for the continuum result (where $\Delta\cc$ is obtained by subtracting (\ref{eq:condensate_free_theory_continuum_vanishing_B}) from (\ref{eq:condensate_free_theory_continuum})) and the three discretizations. 
One observes that the agreement with the continuum condensate is best for naive fermions. In an 
interacting theory, however, one cannot simulate the $\Nf=1$ model with naive fermions by simply 
dividing by the number of doublers, which is the main reason we refrained from using the naive 
discretization in our study. 

The agreement for overlap fermions is very good for weak magnetic fields,
in particular in the regime of $eB/m^2$ we investigate in our simulations. For
stronger magnetic fields the qualitative behavior is still the same as for the 
continuum result, but the quantitative deviation (which appears to be quadratic in
$B$) is substantial. We accredit this
deviation to discretization artifacts, which for massive overlap fermions are 
worse ($\mathcal{O}(a)$) than for naive fermions ($\mathcal{O}(a^2)$). One should
therefore be cautious when interpreting our simulation results -- while we do 
believe in their qualitative correctness, the absolute numbers could be
systematically underestimated at large magnetic fields.
In future studies one could employ an improvement program, such as the one suggested
in \cite{IH09p}, to reduce discretization effects.

For SLAC fermions, perhaps unsurprisingly, the agreement with the continuum result
is rather poor, as the SLAC condensate does not even reproduce the qualitative 
features of the continuum one, for instance, the dip for the lowest allowed 
magnetic field. 
We mention a number of (ultimately futile) attempts we experimented with in order 
to improve the SLAC derivative in a magnetic field given in 
\eqref{eq:slac_magnetic_field}. 

First, we tried out different gauges instead of 
(\ref{eq:symmetric_gauge}), the latter leading to the best agreement, however. Next, we considered a 
physical situation where the magnetic field is constant and positive in one half of the lattice and 
constant and negative (with the same absolute value) in the other half. This avoids the need for 
introducing the lattice boundary terms in \eqref{eq:lattice_vector_potential}
entirely, which for the SLAC 
formulation were quite awkward in the first place. We then only considered the chiral condensate on a 
single lattice point $x$, lying in the center of the region with positive magnetic field. This was 
motivated by the intuition that at such a point the influence from the region with negative magnetic 
field should be negligible for large enough lattices. However, the agreement with continuum results 
we found was still poor. We conclude that more work is necessary if one aims at
making SLAC fermions work for a background magnetic field.

	\section{Proof there is no sign problem}\label{app:sign_problem}
	We show that there is no complex-action problem in the overlap formalism (\ref{eq:overlap}) by 
showing that $\det D$ is real and non-negative. To this end, we work with the following representation 
of gamma matrices:
\begin{align}
	\gamma_\mu = \begin{pmatrix}\sigma_\mu&0 \\0&-\sigma_\mu\end{pmatrix}\;,
\end{align}
where the $\sigma_\mu$ can be chosen as the usual Pauli matrices. This decomposition makes clear how the reducible 
representation we use in this work is made up of the two inequivalent irreducible representations 
in three space-time dimensions, $\sigma_\mu$ and $-\sigma_\mu$.

It is then straightforward to convince oneself that the overlap operator (\ref{eq:overlap}) also 
assumes a block form:
\begin{align}\label{eq:overlap_block_diagonal}
	D = \begin{pmatrix}D_1&0\\0&D_2\end{pmatrix}\;,
\end{align}
where ($i=1,2$)
\begin{align}
	D_i &= D_{\mathrm{ov},i} + \sigma\left(1-\frac{a}{2}D_{\mathrm{ov},i}\right)\;,\\
	D_{\mathrm{ov},i} &= \frac{1}{a}\left(\id+A_i\big/\sqrt{A_i^\dagger A_i}\right)\;,\\
	A_i &= D_{W,i}-\id\;,
\end{align}
and the irreducible components of the Wilson operator read (see \eqref{eq:covariant_difference} for 
the definitions of $\nabla_\mu$ and $\nabla_\mu^*$)
\begin{align}\label{eq:irreducible_wilson_components}
	\begin{aligned}
		D_{W,1} &= \frac{1}{2}\left[\sigma_\mu(\nabla_\mu^*+\nabla_\mu)-a\nabla_\mu^*\nabla_\mu\right]
		\;,
		\\
		D_{W,2} &= \frac{1}{2}\left[-\sigma_\mu(\nabla_\mu^*+\nabla_\mu)-
		a\nabla_\mu^*\nabla_\mu\right]\;.
	\end{aligned}
\end{align}

We emphasize that the diagonal elements $D_{1,2}$ in (\ref{eq:overlap_block_diagonal}) are precisely 
the expressions one would obtain for the overlap operator when working in one of the two irreducible 
representations. Hence, $D$ decomposes in complete analogy to the continuum Dirac operator.

\begin{table*}[t]
	\caption{Parameter sets used in the simulations. Here, $\Ns$ denotes the spatial lattice extent, 
	assumed 
	equal in both directions, $\Nt$ is the temporal extent, $g^2$ denotes the coupling constant in 
	\eqref{eq:lattice_action}, $b$ is the magnetic flux quantum number in 
	\eqref{eq:flux_quantization} and $T_0$ denotes the temperature at which we set the scale 
	$a\sigma_0$ in \eqref{eq:scale}, which we quote in lattice units here. For the 
	scale-setting data, the dots indicate steps of $0.005$. As was explained in 
	\sref{sec:scale_setting}, we use different $T_0$ for the infinite-volume and continuum 
	extrapolations.}
	\renewcommand{\arraystretch}{1.5}
	\renewcommand{\tabcolsep}{10pt}
	\newcommand{\ncol}{6}
	\begin{tabular}{cccccc}
		\hline\hline
		$\Ns$ & $1/g^2$ & $\Nt$ & $b$ & $T_0/\sigma_0$ & $a\sigma_0$ \\
		\hline
		\multicolumn{\ncol}{c}{\textbf{Scale-setting}}\\
		\hline\\[-0.5cm]
		\makecell{$8$ \\ $10$ \\ $12$ \\ $14$ \\ $16$} & 
			\makecell{$0.150$, $\hdots\,$, $0.205$ \\
					  $0.150$, $\hdots\,$, $0.205$ \\
					  $0.150$, $\hdots\,$, $0.190$, $0.200$, $0.205$ \\
					  $0.150$, $\hdots\,$, $0.165$, $0.175$, $\hdots\,$, $0.205$ \\
					  $0.150$, $\hdots\,$, $0.200$} & 
			\makecell{$8$ \\ $10$ \\ $12$ \\ $14$ \\ $16$} & $0$ & $-$ & $-$\\
		\hline
		\multicolumn{\ncol}{c}{\textbf{infinite-volume extrapolation}} \\
		\hline\\[-0.5cm]
		$8$ & $0.1520$ & \makecell{$2$, $3$, $4$, $5$, $6$, $7$, $8$, $12$ \\ $16$} & \makecell{$0$, $1$, 
		$2$, $3$, $4$ \\ $0$} & $0.059$ & $1.063$ \\
		\hline\\[-0.5cm]
		$12$ & $0.1520$ & \makecell{$2$, $4$, $5$, $7$, $8$ \\ $6$ \\ $12$ \\ $16$} & \makecell{$0$, 
		$1$, $3$, $6$, $8$ \\ $0$, $1$, $3$, $5$, $6$, $8$ \\ $0$, $1$, $2$, $3$, $4$, $6$, $8$ \\ 
		$0$} & $0.062$ & $1.004$\\
		\hline
		$16$ & $0.1520$ & $2$, $4$, $5$, $6$, $7$, $8$, $12$, $16$ & $0$, $1$, $2$, $4$, $8$, $12$ & 
		$0.063$ & $0.987$\\
		\hline
		\multicolumn{\ncol}{c}{\textbf{continuum extrapolation}}\\
		\hline
		$8$ & $0.1520$ & $2$, $3$, $4$, $5$, $6$, $7$, $8$ & $0$, $1$, $2$, $3$, $4$ & $0.126$ & 
		$0.995$\\
		\hline\\[-0.5cm]
		$12$ & $0.1650$ & \makecell{$2$, $4$, $5$, $6$, $7$, $8$ \\ $12$} & \makecell{$0$, $1$, $3$, 
		$6$, $8$ \\ $0$, $1$, $2$, $3$, $4$, $6$, $8$} & $0.121$ & $0.691$\\
		\hline\\[-0.5cm]
		$16$ & $0.1740$ & \makecell{$2$, $4$, $6$, $8$, $12$ \\ $16$} & \makecell{$0$, $1$, $3$, $6$, 
		$12$, $14$ \\ $0$, $1$, $2$, $3$, $4$, $6$, $8$, $12$, $14$} & $0.136$ & $0.460$\\
		\hline
		\hline
	\end{tabular}
	\label{tab:parameters}
\end{table*}

Now, obviously,
\begin{align}\label{eq:determinant_product}
	\det D = \det D_1\det D_2\;.
\end{align}
Furthermore, we note that the symmetric difference operator $\nabla_\mu^*+\nabla_\mu$ in 
(\ref{eq:irreducible_wilson_components}) is 
anti-Hermitian, while the discretized Laplacian $\nabla_\mu^*\nabla_\mu$ is Hermitian, such that 
$D_{W,1}$ and $D_{W,2}$ are Hermitian conjugates of one another. By using the spectral representation of the inverse square root 
in the definition of $D_{\mathrm{ov},2}$, one can 
then show that the same holds for $D_1$ and $D_2$, such that, using \eqref{eq:determinant_product},
\begin{align}
	\det D = \det D_1\det D_1^\dagger = \vert\det D_1\vert^2 \geq 0\;,
\end{align}
i.e., there is no complex-action problem since the determinant is real and non-negative. 
We emphasize that the crucial ingredient for this proof was the use of a reducible representation of 
gamma matrices. 
	
	\section{Parameters}
	In our simulations we generated $\mathcal{O}(10^3)-\mathcal{O}(10^4)$ configurations per parameter 
set. We performed binned jackknife resamplings for our error analyses, making sure that each bin 
contained at least $\tau_\mathrm{int}$ configurations (but most commonly multiples thereof), where $
\tau_\mathrm{int}$ refers to the integrated auto-correlation time corresponding to the order parameter 
$\sig$. We found $\tau_\mathrm{int}\lesssim50$ in all cases.

We list the relevant parameters for which we have obtained simulation data, as well as the respective 
scales $\sigma_0$ and scale-setting temperatures $T_0$, in \tref{tab:parameters}. Notice that the 
scale-setting temperatures are different between the infinite-volume and continuum limits, see 
\sref{sec:scale_setting}. Since the errors in $\sigma_0$ are negligible, we do not quote them here and 
refrain from taking their influence on error propagation into account in the entirety of this work.
\label{app:parameters}
	
	\onecolumngrid
	\FloatBarrier	
	\twocolumngrid
	
	\bibliography{bibliography}

\end{document}